\documentclass[aps,prd,superscriptaddress,nofootinbib,11pt]{revtex4}
\usepackage[english]{babel}
\usepackage[utf8]{inputenc}
\usepackage{graphicx}   
\usepackage{slashed}
\usepackage{epstopdf}
\usepackage{verbatim}   
\usepackage{color}      
\usepackage{subfigure}  
\usepackage{multirow}
\usepackage{hyperref}   
\usepackage{float}
\usepackage{epsfig,rotating}
\usepackage{amsmath,amssymb}
\usepackage{dsfont}
\usepackage{slashed}
\restylefloat{table}
\numberwithin{equation}{section}

\newcommand{\vp}{\vec{p}}
\newcommand{\vq}{\vec{q}}
\newcommand{\vk}{\vec{k}}

\newcommand{\be}{\begin{equation}}
\newcommand{\ee}{\end{equation}}
\newcommand{\bea}{\begin{eqnarray}}
\newcommand{\eea}{\end{eqnarray}}

\newcommand{\ket}[1]{|#1\rangle}
\newcommand{\bra}[1]{\langle#1|}

\begin{document}
\title{Threshold and infrared singularities:  time evolution, asymptotic state and entanglement entropy.}

\author{Daniel Boyanovsky}
\email{boyan@pitt.edu} \affiliation{Department of Physics and
Astronomy, University of Pittsburgh, Pittsburgh, PA 15260}

 \date{\today}

\begin{abstract}
 Threshold and infrared divergences are studied as possible mechanisms of particle production and compared to the usual decay process in a model quantum field theory from which generalizations are obtained. A spectral representation of the propagator of the decaying particle suggests that decay, threshold and infrared singularities while seemingly different phenomena are qualitatively related. We implement a non-perturbative dynamical resummation method  to study the time evolution of an initial  state. It is manifestly unitary and yields the asymptotic state and the distribution function of produced particles. Whereas the survival probability in a   decay process falls off as $e^{-\Gamma t}$, for threshold and infrared divergent cases it falls off instead as $e^{-\sqrt{t/t^*}}$ and $t^{-\Delta}$ respectively, with $\Gamma, \Delta \propto (coupling)^2$ whereas $1/t^* \propto (coupling)^4$. Despite the different decay dynamics, the asymptotic state is qualitatively similar: a kinematically entangled state of the daughter particles with a distribution function  which fulfills the unitarity condition and is strongly peaked at energy conserving transitions but broadened by the ``lifetime'' $1/\Gamma~;~ t^*$ for usual decay and threshold singularity, whereas it scales with the anomalous dimension $\Delta$ for the infrared singular case. Threshold and infrared instabilities are production mechanisms just as efficient as particle decay. If one of the particles is in a dark sector and  not observed, the loss of information yields an entanglement entropy  determined by the distribution functions and increases upon unitary time evolution.

\end{abstract}


\maketitle

\section{Introduction}

Most particles in the standard model decay,   quarks and gluons are confined, and charged particles interacting with gauge fields are dressed by a cloud of soft massless gauge fields. Therefore, of all the particles in the standard model perhaps only neutrinos and photons appear as asymptotic single particle states in the S-matrix. The dressing of charged particles by massless gauge bosons results in infrared divergences in radiative corrections as a consequence of the emission and absorption of the soft gauge quanta. Understanding these infrared phenomena and the infrared finiteness of the S-matrix has been\cite{bn,lee,chung,kino,kibble,yennie,weinberg,kulish,greco} and continues\cite{lavelle,ein,noji,haya,carney,zell,tomaras,neli,noji} to be the focus of a substantial body of work motivated by precision calculations of physical observables for collider experiments\cite{schwartz1,finites,schwartz2}. Infrared phenomena also plays a fundamental role in quantum aspects of gravity as a consequence of emission and absorption of gravitons\cite{strominger1,strominger2}.

Prior to the discovery of the Higgs boson, early work\cite{kni,will} recognized that the S-matrix approach to describing particle decay breaks down when the mass of the particle approaches the multiparticle threshold\cite{kni,will,thres1,thres}. In particular refs.\cite{kni,will,thres1,thres} recognized a singularity in the self-energy of the   particle as its mass approaches threshold from below, and as a consequence the particle no longer appears as an asymptotic state in the S-matrix.

Notably this situation is  similar to the case of infrared singularities in gauge theories that arise because the mass of the charged particle coincides with the multiparticle threshold suggesting that, perhaps,   threshold and infrared singularities, although quantitatively different,  are manifestations  of similar phenomena suggesting a generalized decay of the particle.

\vspace{1mm}

\textbf{Motivations and objectives:}

Extensions beyond the standard model posit the existence of new particles as possible explanations of the origin of  dark matter in cosmology. Some of these extensions introduce light or ultralight particles\cite{axionreviu,axionsikivie,abbott,dine,fuzzyDM,fuzzy2,wittenfuzzy}, and an important question in these models is to identify and assess the production mechanism for these dark matter candidates. A recent study\cite{infra}  revealed certain universality of infrared phenomena in the sense that infrared divergences associated with emission and absorption of massless quanta feature similar dynamics and asymptotic states in bosonic, fermionic and (abelian) gauge theories. This study also revealed that the infrared divergences could be an effective production mechanism of soft massless particles, and was extrapolated to the realm of production of light dark matter or dark radiation during a radiation dominated cosmology\cite{infrafrw}. In this article we extend the study of ref.\cite{infra} to compare and contrast the dynamics of decay, threshold and infrared divergences to identify hitherto unexplored production mechanisms that could be relevant in early Universe cosmology and also, perhaps, of some phenomenological interest in particle physics.

  As windows beyond the standard model open to explore possible explanations of dark matter and or dark radiation, our study is  motivated by its possible impact in identifying and assessing alternative production mechanisms available in the dark sector, but also to explore fundamental aspects of the dynamics of particle decay, threshold and infrared divergences that could be of  a more overarching phenomenological and theoretical interest.

\textbf{Objectives:}  Our objectives in this study are the following: i) to compare and contrast the dynamical aspects of particle decay and threshold and infrared divergences within a model quantum field theory and draw more general conclusions on the time evolution of initial towards asymptotic states, ii) to understand threshold and infrared singularities as possible production mechanisms and to explore a qualitative similarity between      these seemingly different phenomena, iii) to understand     the time evolution that leads from the initial to the final asymptotic state and to characterize the properties of the latter,
iv) for threshold and infrared divergences the usual decay rates vanish, therefore understanding the time evolution of initial states   will clarify the dynamics of relaxation towards equilibrium in these cases.

Our study  does \emph{not} address the important issues of the infrared finiteness of the S-matrix, a far broader subject of much current interest\cite{schwartz1,finites,schwartz2,neli}. It is much more narrowly focused on understanding the time evolution of states and the emerging  asymptotic states in the case of threshold and infrared divergences.
A  reassessment\cite{collins}  of the Lehmann, Symanzik and Zimmermann reduction formula for asymptotic states beginning with a finite time analysis and extending it  to the infinite time limit has highlighted the subtleties of this limit.

Our study in this article may provide complementary further insights into asymptotic theory in cases in which threshold and infrared divergences substantially modify the asymptotic long time dynamics, and may contribute to  the fundamental  understanding of the asymptotic states emerging from these processes.

\vspace{1mm}

\textbf{Brief summary of results:}
We study decay, threshold and infrared phenomena within a simple model of a real scalar field $\Phi$ coupled to two other scalar fields of different masses  that effectively captures the different phenomena by varying the various masses. The Kallen-Lehmann representation of the propagator of the $\Phi$ field including radiative corrections illustrates how decay, threshold and infrared phenomena, although seemingly disparate are qualitatively related. Furthermore, it clearly shows the breakdown of a Breit-Wigner approximation as the mass of the particle approaches threshold.

 A dynamical resummation method\cite{dbzeno,infra} is implemented to study the time evolution of an initial single particle state of the $\Phi$ field  towards the final asymptotic state in all cases. This method is manifestly unitary and complementary to the dynamical renormalization group\cite{gold,boyvega}. It yields not only the time evolution of the initial state, but also describes the emergence of the asymptotic state during the evolution and its properties.

We find that whereas the time evolution of the survival probability of a single particle state in a typical decay process is $e^{-\Gamma t}$, in the cases of threshold and infrared singularities the usual decay rate vanishes, however we find that the survival probability of the initial state indeed decays: in the case of threshold divergence it evolves as $e^{-\sqrt{t/t^*}}$ and for infrared divergences as $\propto t^{-\Delta}$. Whereas $\Gamma$ and the anomalous dimension $\Delta$ are of $\mathcal{O}(g^2)$ with  $g$ the coupling, the relaxation time scale $t^* \propto 1/g^4$ as a consequence of the threshold singularity.

We find that despite the different time evolution, the asymptotic state is qualitatively similar: a  kinematically entangled state of the daughter particles with pair correlations.    We obtain the probabilities of these pairs, show that they satisfy the unitarity condition and identify them as the \emph{distribution function} of the produced particles which are obtained in each case. Although these are peaked at energy conserving transitions, are much narrower in the case of threshold  divergences as a consequence of a longer ``lifetime'' of the initial state and feature a scaling behavior with the  anomalous dimension $\Delta$ in the case of infrared divergences.

A corollary of this result is that threshold and infrared singularities are just as efficient production mechanisms as decay.

These asymptotic states are very different from those postulated in quantum electrodynamics\cite{kibble,chung,kulish,tomaras} as solutions to the infrared problem, but are unambiguously obtained from the unitary time evolution of an initial state. We argue that the pair correlations in the asymptotic state, in other words the entanglement of the daughter particles, implies the same distribution function for each, which we obtain from the time evolution in all cases. If either one of the daughter particles is not measured for example in the ``invisible decay'' into a dark matter particle, the information loss leads to an entanglement entropy, which is shown to grow during the time evolution from the initial to the asymptotic state.

\section{ Kallen-Lehmann spectral representation:}\label{sec:specrep}
    We consider a model of a massive real  scalar field $\Phi$ coupled to two other real scalar fields $\chi_1,\chi_2$, to illustrate the main phenomena within a simpler setting, with the objective of drawing more general conclusions. Such model has previously been investigated within the context of threshold singularities in refs.\cite{will,thres}. It is described by the following Lagrangian density

\be
\mathcal{L} = \frac{1}{2}\,\partial^\mu \Phi  \partial_\mu \Phi - \frac{1}{2}\,M^2 \Phi^2 +
\frac{1}{2}\,\partial^\mu \chi_1\,  \partial_\mu \chi_1- \frac{1}{2} \,m^2_1 \,\chi^2_1 +\frac{1}{2}\,\partial^\mu \chi_2\,  \partial_\mu \chi_2- \frac{1}{2} \, m^2_2 \,\chi^2_2\,-
\lambda \,\Phi     \chi_1 \chi_2 \,. \label{lagsuper}\ee

\begin{figure}[ht!]
\begin{center}
\includegraphics[height=2in,width=2in,keepaspectratio=true]{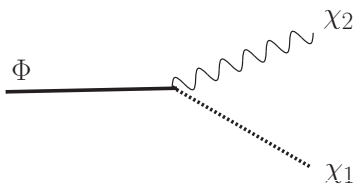}
\caption{Interaction vertex in the theory defined by the Lagrangian density (\ref{lagsuper}).}
\label{fig1:vertex}
\end{center}
\end{figure}

\begin{figure}[ht!]
\begin{center}
\includegraphics[height=2.5in,width=2.5in,keepaspectratio=true]{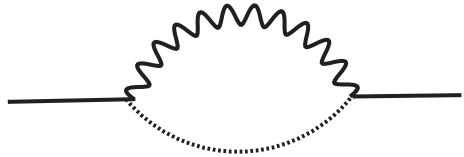}
\caption{One loop self energy of the $\Phi$ field in the theory defined by the Lagrangian density (\ref{charged}).}
\label{fig2:selfenergy}
\end{center}
\end{figure}

This Lagrangian density provides a simple arena to study the main aspects of our focus in this article: i) if $M < (m_1+m_2)$, a single $\Phi$ particle is stable, ii) when $M> (m_1+m_2)$ a $\Phi$ particle is unstable and decays into a pair of $\chi_1,\chi_2$ particles, iii) when $M=(m_1+m_2)$ the mass of the $\Phi$ particle is exactly at threshold and this case is a manifestation of the threshold singularity, studied originally in ref.\cite{will}, iv) infrared singularity when $M=m_1~,~ m_2=0$ arising from the emission and absorption of massless quanta. In this case again the mass of the particle $\Phi$ coincides with the multiparticle threshold. This latter case features the same infrared singularities as that of a charged field coupled to a massless field studied in ref.\cite{infra}   within a bosonic model with Lagrangian density
\be \mathcal{L} =  \partial^\mu \Phi^\dagger  \partial_\mu \Phi -  M^2 \Phi^\dagger \Phi +
\frac{1}{2}\,\partial^\mu \chi\,  \partial_\mu \chi-
\lambda \,\Phi^\dagger \Phi  \, \chi  \,. \label{charged}\ee In this model the infrared singularity emerges in the self-energy of the $\Phi$ field as a consequence of the emission and absorption of massless quanta $\Phi \leftrightarrow \Phi \chi$. In ref.\cite{infra} it is shown that the infrared behavior of this model is similar to that of a Dirac fermion Yukawa coupled to a massless scalar (a  renormalizable theory), and in turn is similar to the infrared divergence of the fermionic self-energy in quantum electrodynamics.  Hence, the Lagrangian (\ref{lagsuper}) furnishes a simple   quantum field theory that allows to study all four cases: i) stable, ii) unstable, iii) threshold singularity and iv) infrared divergence within the same model by adjusting the masses  appropriately.  Fig. (\ref{fig1:vertex}) depicts the interaction vertex in the theory described by (\ref{lagsuper}) and fig. (\ref{fig2:selfenergy}) shows the one-loop self energy of the field $\Phi$ in the theory described by (\ref{charged}) which features an infrared divergence, this self energy is the same as that obtained from (\ref{lagsuper}) replacing $\chi_1 \rightarrow \Phi~~;~~ \chi_2 \rightarrow \chi$.

The Lagrangian density (\ref{lagsuper}) describes a super renormalizable theory, however, because  we are interested in infrared and long time phenomena which we expect to be insensitive to the ultraviolet behavior of the theory, this model is expected to capture the long time dynamics reliably. This expectation is confirmed by the study of ref.\cite{infra} where infrared phenomena and long time dynamics were shown to be the same for a superrenormalizable and a renormalizable model. Furthermore, in ref.\cite{dbzeno} it has been shown that ultraviolet divergences contribute to very early transients that do not affect the long time dynamics and can be safely absorbed into a renormalization of the initial amplitude. This is a consequence of the wide separation of time scales between the  ultraviolet early transients and the long time infrared phenomena. Taken together the results of these previous studies serve as anchors that   allow us to draw more general conclusions on the long time dynamics from the simple model described by eqn. (\ref{lagsuper}).

We begin by studying the Kallen-Lehmann spectral representation\cite{barton} of the single $\Phi$ particle propagator including a Dyson resummation of the one loop self-energy shown in fig. (\ref{fig2:selfenergy}). The propagator is given by
\be G(P^2) = \frac{1}{P^2 - M^2 - \Sigma(P^2)+i\epsilon} \,,\label{fiprop} \ee the self-energy is calculated in dimensional regularization in dimension $D= 4-\varepsilon$, and introducing a renormalization scale $\mu$ we find
\be \Sigma(P^2) = -\frac{\widetilde{\lambda}^2}{(4\pi)^2}\,L + \frac{\widetilde{\lambda}^2}{(4\pi)^2}\,I(P^2/M^2)\,, \label{sigsup1} \ee
  where
\be \widetilde{\lambda} = \lambda\,\mu^{-\varepsilon/2}~~;~~ L = \frac{2}{\varepsilon}-\gamma_E + \ln(4\pi) - \ln\Big[\frac{M^2}{\mu^2} \Big] \label{msbar} \ee with $\gamma_E$ the Euler-Mascheroni constant and
\be I(P^2) = \int^1_0 \ln\Big[\frac{m^2_2}{M^2}+\frac{m^2_1-m^2_2}{M^2}\,x-\frac{P^2}{M^2}\,x\,(1-x)-i\tilde{\epsilon} \Big]\, dx ~~;~~ \tilde{\epsilon} \rightarrow 0^+ \,. \label{Iofalfa} \ee

Separating the real and imaginary parts of the self-energy,
\be \Sigma(P^2) =  \Sigma_R(P^2) + i \,\Sigma_I(P^2)\,, \label{sepsig}\ee we find
\be  \Sigma_R(P^2) = -\frac{\widetilde{\lambda}^2}{(4\pi)^2}\,L + \frac{\widetilde{\lambda}^2}{(4\pi)^2}\,
\int^1_0 \ln \Bigg|\frac{m^2_2}{M^2}+\frac{m^2_1-m^2_2}{M^2}\,x-\frac{P^2}{M^2}\,x\,(1-x) \Bigg|  \, dx
\label{Resigma}\ee

\be  \Sigma_I(P^2)= - \pi\,\frac{\widetilde{\lambda}^2}{(4\pi)^2}\,\Bigg[\Big(1-\frac{(m_1+m_2)^2}{P^2}\Big)\Big(1-\frac{(m_1-m_2)^2}{P^2}\Big) \Bigg]^{1/2}\,\Theta\Big(P^2-(m_1+m_2)^2\Big)\,.\label{Imsigma} \ee

Subtracting the real part of the self-energy at $P^2 = M^2_p$   at which the real part of the inverse propagator vanishes, namely
\be  \Sigma_R(P^2) =  \Sigma_R(P^2=M^2_p) +  \widetilde{ \Sigma}_R(P^2)\,, \label{subt}\ee where
\be M^2_p = M^2 +  \Sigma_R(P^2=M^2_p)\,, \label{pole} \ee and to leading order in the coupling replacing $M\rightarrow M_p$ in the expression  for the real and imaginary  parts of the self-energy (\ref{Resigma}), it follows that
\be G(P^2) = \frac{1}{P^2-M^2_p -  \widetilde{\Sigma}_R(P^2)-i \Sigma_I(P^2) + i\epsilon} \,. \label{renprop}\ee

The Kallen-Lehmann spectral function is given by \cite{barton}
\be \sigma(P^2) = -\frac{1}{\pi} \,\mathrm{Im}\, G(P^2) = \frac{1}{\pi}\,\frac{-\Sigma_I(P^2)+\epsilon}{\Big[P^2-M^2_p -  \widetilde{\Sigma}_R(P^2) \Big]^2+\Big[-\Sigma_I(P^2)+\epsilon \Big]^2}\,,\label{specfun}\ee it contains the information on the asymptotic properties of the quanta of the real scalar field $\Phi$ and obeys the sum rule
\be \int \sigma(P^2) \, dP^2 =1 \,.\label{sumrule} \ee

 A single particle pole below threshold, namely for  $P^2 <(m_1+m_2)^2$, for which $\Sigma_I(P^2) =0$,  yields
  \be \sigma_p(P^2) = Z \, \delta(P^2-M^2_p)  \,,\label{singlepar} \ee where
   \be Z^{-1} = 1-\frac{\partial \,\widetilde{\Sigma}_R(P^2)}{\partial P^2}\Big|_{P^2=M^2_p}\,. \label{Zdef} \ee

   The wave function renormalization constant yields the amplitude of the single particle pole and determines the overlap
   between the bare single particle state and the asymptotic renormalized state of a \emph{stable} particle that has been dressed by quantum fluctuations. Therefore, when the single particle pole is below threshold, the particle is stable and
   \be \sigma(P^2) = Z \, \delta(P^2-M^2_p) + \sigma_{c}(P^2) \,, \label{stable} \ee where $\sigma_c(P^2)$ is the contribution from the multiparticle continuum above threshold. In this case   the sum rule (\ref{sumrule}) yields
   \be Z + \int_{P^2_T} \sigma(P^2) \, dP^2 =1 ~~;~~ P^2_T = (m_1+m_2)^2\,, \label{belth} \ee whereas if the particle decays and the single particle pole is embedded in the continuum, there is no single particle pole below threshold and the sum rule (\ref{sumrule}) yields
   \be \int^{\infty}_{P^2_T} \sigma(P^2) \, dP^2 =1\,, \label{aboth} \ee namely it is saturated by the continuum ``background'' and the single particle quanta of the $\Phi$ field are not   asymptotic states.

   \section{Decay, threshold and infrared singularities:}\label{sec:dethir}

   \subsection{Decay and threshold singularities}\label{subsec:deth}

   In order to discuss both cases of decay and threshold singularities we consider the simpler case of equal masses $m_1=m_2 \equiv m$ when the two particle threshold is at $P^2 = 4m^2$, furthermore, it is convenient to introduce the dimensionless variables
   \be s = \frac{P^2}{M^2_p} ~~;~~ r = \frac{4m^2}{M^2_p}~~;~~g= \Big(\frac{\widetilde{\lambda}}{4\pi M_p}\Big)^2~~;~~\Delta(s,r)= \sqrt{1-\frac{r}{s}}\,.  \label{dimratios}\ee In terms of these variables
   the spectral density becomes
   \be M^2_p \sigma(s,r)=
   \frac{ g \,\Delta(s,r)\,\Theta(s-r)+\epsilon}{\Big[s-1-g\,D(s,r)\Big]^2+\Big[\pi\,g\,\Delta(r,s)\,\Theta(s-r)+\epsilon \Big]^2  }~~;~~ \epsilon \rightarrow 0^+ \,,\label{sigfin}\ee
   where
   \be D(s,r)= \Delta(s,r) \ln\Bigg[\frac{1+\Delta(s,r) }{1-\Delta(s,r)}\Bigg]- \Delta(1,r) \ln\Bigg[\frac{1+\Delta(1,r) }{1-\Delta(1,r)}\Bigg]\,. \label{Dsr}\ee

   We study the cases $r>1$ (stable particle) and $r<1$ (unstable decaying particle) separately to highlight both differences and similarities.

  \vspace{1mm}

  \textbf{ Case I: Stable particle  $4m^2 > M^2_p$ ($r>1$)}

\begin{figure}[h!]
\begin{center}
\includegraphics[height=4in,width=4in,keepaspectratio=true]{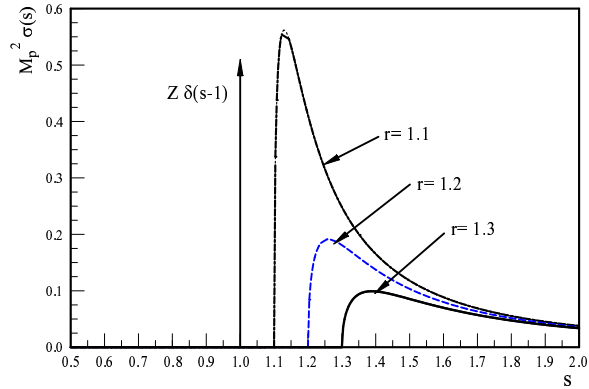}
\caption{$M^2_p\,\sigma(s)$ for $r= 1.1, 1.2, 1.3 $ and $g=0.01$ describing a stable particle with an isolated pole below threshold. $s= P^2/M^2_P~~;~~r=4m^2/M^2_p$. }
\label{fig3:specabove}
\end{center}
\end{figure}

  In this case   the propagator features  an isolated single particle pole at $s=1$ below the two particle threshold at $r>1$
  and for $s<r$
   \be M^2_P \, \sigma(s,r)=  Z(r)\,\delta(s-1)~~;~~ s<r  \label{isopole} \ee with $Z^{-1}$ given by eqn. (\ref{Zdef}) for which we find
   \be Z(r) = \frac{1}{ [1+ g\, \Big[\Big(\frac{1}{\overline{\delta}(r)}+\overline{\delta}(r) \Big)\,atan\Big(\frac{1}{\overline{\delta}(r)} \Big)-1 \Big]  } ~~;~~ \overline{\delta}(r) = \sqrt{r-1} \,.\label{Zeta}\ee The full spectral density in this case when the particle pole is below the two particle threshold is given by
    \be M^2_p \, \sigma(s,r)=  Z(r) \,\delta(s-1) + \frac{ g \,\Delta(s,r)\,\Theta(s-r)}{\Big[s-1-g\,D(s,r)\Big]^2+\Big[\pi\,g\,\Delta(r,s)\,\Theta(s-r) \Big]^2  } \,. \label{fulsigbelow}\ee
     For $r>1$ the particle is present as an asymptotic state with probability $Z(r) < 1$. However, we note that as $ M^2_p \rightarrow 4 m^2$, namely as the position of the single particle pole approaches the threshold from below, or $r\rightarrow 1$ from above,  the residue at the isolated pole below threshold \emph{vanishes} as
    \be Z(r) ~~~ {}_{ \overrightarrow{r\rightarrow 1}} ~~~
     \frac{1}{1+   \frac{g\, \pi}{\sqrt{r-1}}} \simeq \frac{\sqrt{\frac{4m^2}{M^2_p}-1}}{\pi\,g}\,, \label{vanishzeta}\ee
       with a square root singularity, and very sharply in weak coupling, obviously this behavior is strongly non-perturbative. Furthermore, we find that   while the continuum contribution to the spectral density vanishes at threshold, it becomes sharply peaked near threshold as $r \rightarrow 1$ from above (or $M^2_p \rightarrow 4 m^2$ from below).      Fig. (\ref{fig3:specabove}) displays the spectral density for $r>1$, namely the case of a stable particle described by
     an isolated pole below threshold.

\begin{figure}[ht!]
\begin{center}
\includegraphics[height=4in,width=4in,keepaspectratio=true]{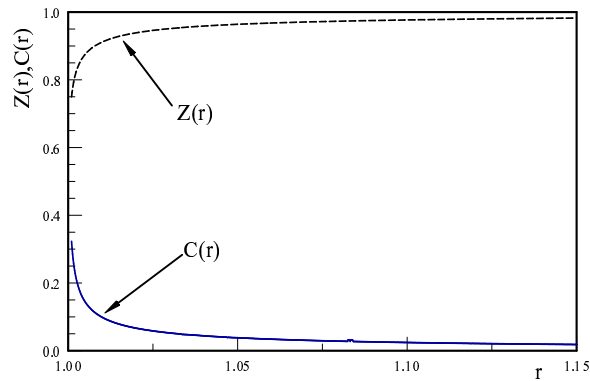}
\caption{Z(r) and C(r) vs.  $r=4m^2/M^2_p$, for $g= 0.01$. The sum rule $Z(r)+C(r)=1$ is confirmed numerically. }
\label{fig4:zcabove}
\end{center}
\end{figure}

  Defining the contribution from the two particle continuum  above threshold as
  \be C(r) = \int^{\infty}_{r}  M^2_p \,\sigma(s,r)\,ds \label{conti}\ee we have confirmed numerically that the sum rule (\ref{belth})
  \be Z(r)+ C(r) =1 \label{sumita}\ee is fulfilled. As $r\rightarrow 1$ from above, the residue at the pole vanishes, but the continuum contribution saturates the sum rule. Fig. (\ref{fig4:zcabove})   shows Z(r) and C(r), it clearly displays that  Z(r) vanishes sharply and C(r) rises sharply  as $M^2_p \rightarrow  4m^2$ from below ($r\rightarrow 1^+$), in agreement with the sum rule (\ref{conti}) which can be confirmed from the figure.

Precisely at $M^2_p = 4m^2$ when the mass shell coincides with the multiparticle threshold there is a singularity in the sense that the amplitude of the single particle pole vanishes and the spectral density at $P^2 = 4m^2$ diverges in such a way as to maintain the sum rule. This behaviour has been described as a threshold singularity\cite{will}. What is clear in the case when $M^2_p= 4m^2$ is that the single particle ``dissolves'' into the continuum and is not an asymptotic state since its residue, namely the overlap of the bare and asymptotic state vanishes. However, the particle does not ``decay'' in the usual manner because the imaginary part of the self-energy vanishes at $P^2 = M^2_p = 4m^2$, hence the ``decay rate'' $\Gamma \propto \Sigma_I(P^2=M^2_p)/M_p $ vanishes identically when $M^2_p=4m^2$.

\vspace{1mm}

\textbf{Case II: Unstable particle  $M^2_p > 4m^2$ ($r<1$)}

\begin{figure}[h!]
\begin{center}
\includegraphics[height=4.5in,width=4.5in,keepaspectratio=true]{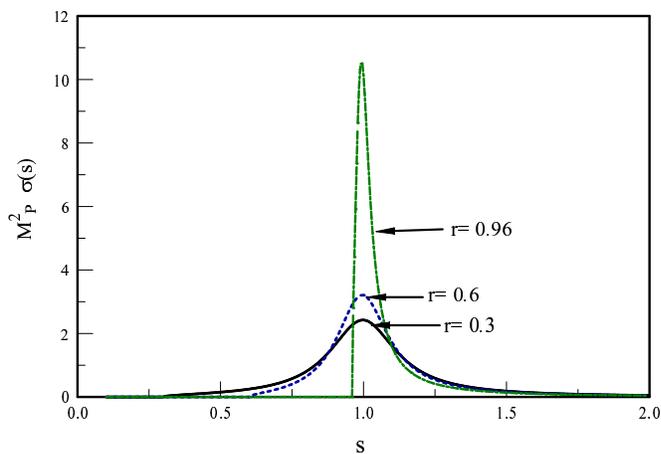}
\caption{$M^2_P\,\sigma(s) \, vs. \, s=P^2/M^2_p$, for an unstable, decaying particle with $M^2_P > 4m^2$ ($r<1$) for $r=0.3,0.6,0.96;g= 0.05$.   }
\label{fig5:sigbelow}
\end{center}
\end{figure}

In this case the particle ``pole'' moves off the physical sheet into the second (or higher) Riemann sheet becoming a decaying resonant state which is not an asymptotic state in the S-matrix.  The spectral density   only has
support above the two particle threshold

 \be M^2_p \, \sigma(s,r)=    \frac{ g \,\Delta(s,r)\,\Theta(s-r)}{\Big[s-1-g\,D(s,r)\Big]^2+\Big[\pi\,g\,\Delta(r,s)\,\Theta(s-r) \Big]^2  } \,. \label{fulsigabove}\ee where $D(s,r)$ is given by eqn. (\ref{Dsr}). It  is displayed in fig. (\ref{fig5:sigbelow}) for $r=0.3,0.6,0.96$ for $g=0.05$, a moderately large coupling to exhibit the behavior as the position of the resonance approaches the threshold from above as compared with the cases where it is far above threshold.

In this case the sum rule (\ref{sumita}) is saturated by the contribution above threshold since there is no support below threshold, and we have confirmed numerically in all cases that $C(r)=1$ with $C(r)$ given by eqn. (\ref{conti}).

When the distance between threshold and the position of the resonance (``pole'') is much larger than the width $\Gamma$ the propagator and the  spectral density may be very well approximated by a Breit-Wigner Lorentzian function in the narrow width approximation,

 \be M^2_p \, \sigma_{bw}(s,r)=   \frac{Z_{bw} }{\pi}\, \frac{ \gamma}{\big[s-1]^2+ \gamma^2  } \,, \label{BWsig}\ee with
 \be \gamma = g\,\pi\,Z_{bw}  \,\Delta(1,r)\,,  \label{gama1}\ee where the wave function renormalization $Z_{bw} $ is given by (\ref{Zdef}) but now above threshold, with $M^2_P > 4m^2$ and given by
 \be Z^{-1}_{bw} = 1-g\,r \Bigg\{\frac{1}{2\Delta(1,r)}\,\ln\Big[ \frac{1+\Delta(1,r)}{1-\Delta(1,r)}\Big]+ \frac{1}{1-\Delta^2(1,r)} \Bigg\}\,,\label{zabove}\ee we note that in contrast to the case when $M^2_P < 4m^2$, in this case as $M^2_P \rightarrow 4 m^2$ from above it is straightforward to confirm that $Z_{bw}$ remains finite in agreement with the conclusion in ref.\cite{will}. However, when the particle is unstable $Z_{bw}$ does not have the interpretation of the amplitude of the renormalized single particle state in the asymptotic state.  It is clear from eqn. (\ref{BWsig}) that the Breit-Wigner approximation of the spectral density is only reliable for very weak coupling as it does not obey the sum rule $C(r)=1$ since $Z_{bw} \neq 1$.

  When the position of the resonance (``pole'') is far away from threshold and for a narrow width, the propagator may be approximated   by a Breit-Wigner distribution which in the narrow width approximation   becomes
 \be G(P^2) = \frac{Z_{bw}}{\Big[P^2-M^2_p  +i M_p \,  \Gamma \Big]}\,, \label{bwgofP}\ee with
 $Z_{bw}$ given by eqn. (\ref{zabove}) and
 \be   {\Gamma}  = - Z_{bw}\,\frac{\Sigma_I(P^2=M^2_p)}{M_p}\,,  \label{gamita}\ee to leading order in the weak coupling $g$, we can set $Z_{bw}=1$   and recognize $\Gamma$ as the decay rate at rest obtained from the lowest order S-matrix approach, namely
 \be \Gamma = \pi\,g\,M_p\,\sqrt{1-\frac{4m^2}{M^2_p}}\,. \label{gammapole}\ee

The long time dynamics of the retarded propagator   is obtained from the Fourier transform of the Breit-Wigner propagator (\ref{bwgofP}), namely
\be G_{ret}(t) = i \int^{\infty}_{-\infty} \frac{dp_0}{2\pi} ~ G(P)\,  e^{-i p_0 t} ~~;~~ t>0 \,,\label{greta}\ee
yielding
\be G_{ret}(t) = Z_{bw}\, \frac{e^{-iE_p\,t}}{2E_p}\, e^{-\Gamma_p\,t/2} ~~;~~ E_p=\sqrt{\vec{p}^2+M^2_p}~~;~~ \Gamma_p = \frac{M_p}{E_p}~\Gamma \label{expdecay} \ee from which we interpret $Z_{bw}$ not as the amplitude of the single particle in the asymptotic state but as the weight of the resonance contribution to the spectral density as evident from eqn. (\ref{BWsig}).

 As fig.(\ref{fig5:sigbelow}) clearly shows, when  $M^2_p \rightarrow 4 m^2$ from above ($r\rightarrow 1$ from below), although
  $\Gamma \rightarrow 0 $, the spectral density can no longer be described as a narrow width Breit-Wigner Lorentzian and the resonance cannot be described as a complex pole in the second (or higher) Riemann sheet. In this limit we find that  as $s \rightarrow 1^+$ and for weak coupling
 \be M^2_P \, \sigma(s,1) \simeq  \frac{1}{g\,\pi^2\,\sqrt{s-1} } \label{nearth} \ee displaying a square root singularity at threshold
 in agreement with fig. (\ref{fig5:sigbelow}). In this case the narrow width Breit-Wigner approximation is neither valid nor useful to describe the resonance near threshold as the spectral density diverges as the threshold is approached. However, this singularity notwithstanding, we confirmed numerically the sum rule $C(1)=1$, but the interpretation of a finite  $Z_{bw}$ as a wave function renormalization associated with the resonance is no longer useful as a description of the asymptotic state.

 As $M^2_p \rightarrow 4m^2$ from below the single particle state is no longer an asymptotic state, however its amplitude does not decay in time  with the usual exponential decay law because the decay rate $\Gamma  \rightarrow 0$ when the ``pole'' coincides with the two particle threshold. Furthermore, as is clear from fig. (\ref{fig5:sigbelow}) and from eqn. (\ref{nearth}) the Breit-Wigner approximation breaks down as $M_p \rightarrow 4m^2$ from above and the time evolution of the resonant state is not an exponential as in the case (\ref{expdecay}) but a more complicated function determined by the square root branch cut beginning at threshold. This time evolution will be studied in detail in the next section.

\vspace{1mm}

\subsection{Infrared singularity:}\label{subsec:IR}

The infrared singularity is associated with the emission and absorption of a massless particle by a massive one. Such is the case, for example, in quantum electrodynamics where the one loop fermion self energy features an infared divergence on the fermion mass shell. Whereas in gauge theories care must be taken to maintain gauge invariance and satisfy Ward identities, the simpler model of a charged scalar field in interaction with a neutral massless field, described by the Lagrangian density (\ref{charged}) features the same  infrared divergence\cite{infra}. In turn, we can study the infrared divergence within the framework of the model described by (\ref{lagsuper}) by taking $m_1=M, m_2=0$.
However, in order to display the emergence of the infrared singularity more clearly, let us consider the case $m_2=0~;~ m_1=m$ and we will explore  the limit $M_p \rightarrow m$ where the infrared divergence becomes  manifest.

\begin{figure}[h!]
\begin{center}
\includegraphics[height=4in,width=4in,keepaspectratio=true]{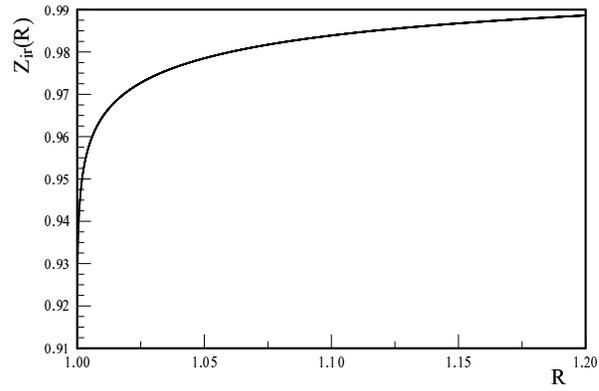}
\caption{$Z_{ir}(R) \, vs. \, R$, for the    infrared case for  $R=m/M_p$ and $g= 0.01$.   }
\label{fig6:zetair}
\end{center}
\end{figure}

In  this case, in terms of the variables $s;g$ introduced in eqn. (\ref{dimratios}) along with the ratio
\be R= \frac{m}{M_p}\,, \label{nuratio}\ee   the spectral density is given by
\be M^2_p \, \sigma(s) = \frac{g\,(\frac{s-R}{s})\,\Theta(s-R)+\epsilon}{\Big[s-1-I(s,R) \Big]^2+\Big[g\,\pi\,(\frac{s-R}{s})\,\Theta(s-R)+\epsilon \Big]^2}~~;~~ \epsilon \rightarrow 0^+\,,
\label{IRsigma}\ee where
\be I(s,R) = g\,\Bigg\{\Big(\frac{s-R}{s}\Big)\,\ln\Big|\frac{s-R}{R} \Big| - (1-R)\,\ln\Big|\frac{1-R}{R} \Big| \Bigg\}\,. \label{IRse}\ee For $R>1$  the $\Phi$ particle is stable and we find
\be M^2_p \, \sigma(s) = Z_{ir}(R)\,\delta(s-1)+M^2_p\,\sigma_c(s)\,, \label{poleIR} \ee
where
\be Z_{ir}(R)= \frac{1}{1+g\,R\,\Big[\ln\big(\frac{R}{R-1}\big)-\frac{1}{R} \Big]}\,, \label{ZIR} \ee and $M^2_p\,\sigma_c(s)$ is the contribution from the two particle continuum above threshold, given by eqn. (\ref{IRsigma}) for  $s>R$. Equation (\ref{ZIR}) clearly shows that as $R\rightarrow 1^+$ the wave function renormalization vanishes, namely there is no longer an isolated single particle pole, again the particle ``dissolves'' into the continuum as its (renormalized) mass approaches the threshold from below.   This behavior is displayed in Fig. (\ref{fig6:zetair}) which shows  $Z_{ir}(R)$ vs. $R$ for $g=0.01$.

\begin{figure}[h!]
\begin{center}
\includegraphics[height=4in,width=4in,keepaspectratio=true]{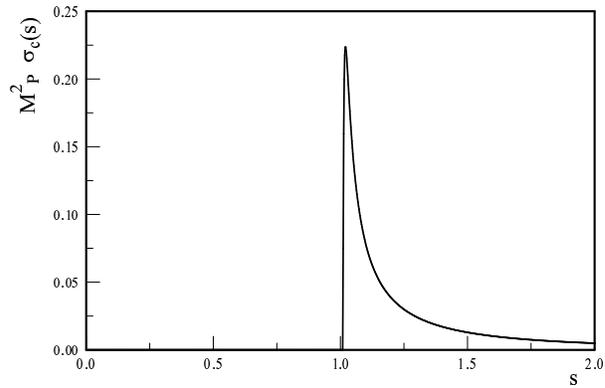}
\caption{$M^2_P\,\sigma_c(s) \, vs. \, s$, for the  the infrared case for  $R=m/M_p=1.01$ and $g= 0.01$.   }
\label{fig7:sigir}
\end{center}
\end{figure}

Again we have confirmed numerically the validity of the sum rule $Z_{ir}(R)+\int^{\infty}_R M^2_p \,\sigma_c(s)ds =1$, therefore when
the single particle pole approaches the threshold from below, the sum rule is saturated from the continuum contribution  $M^2_p\,\sigma_c(s)$ which is displayed in fig. (\ref{fig7:sigir}) showing a  sharp rise near threshold as it absorbs the normalization of the single particle pole when it merges with threshold.

We conclude that in the infrared limit $R=1$ when the single particle pole merges with the threshold the single particle ``dissolves'' into the continuum and is no longer an asymptotic state. However, as in the case of threshold singularity, the particle does not ``decay'' in the usual sense because the decay rate vanishes when the pole mass coincides with the two particle threshold.

The infrared singularity for the case $R=1$ is the same as that studied in the model of a charged scalar field coupled to a massless real scalar field\cite{infra}, this previous study also revealed an emerging universality of infrared phenomena and showed that the amplitude of an initial single particle state decays with a power law with anomalous dimension. The infrared singularities in this model field theory are the same as in the general Lagrangian density (\ref{lagsuper}),   replacing $\chi_1 \rightarrow \Phi~;~ \chi_2 \rightarrow \chi$ and  $m_1=M, m_2 =0$.

The lesson that we draw from this analysis based on the Kallen-Lehmann representation is that threshold and infrared divergences result in that the probability of the single particle state vanishes, transferring the normalization to the multiparticle continuum. This ``flow'' of probability from single particle to multiparticle states is a manifestation of particle production, we refer to these cases as generalized decay, because, indeed the single $\Phi$ particle does decay into the multiparticle continuum despite the fact that the S-matrix decay rate $\Gamma$  formally vanishes, because the imaginary part of the self energy vanishes at threshold.  Furthermore, although there are quantitative differences between threshold and infrared divergences, for example in the manner that $Z$ vanishes as the threshold is approached and the sharp rise of the continuum contribution near threshold, qualitatively the two phenomena are rather similar as evidenced by the figures displaying $Z$ and $\sigma_c$ in both cases.

 In the next section we study this ``flow'' or generalized decay from the point of view of  the time evolution of an initial single particle state towards an asymptotic state.

 \section{\label{sec:WW} Time evolution: dynamical resummation method :}
 We now obtain  the asymptotic state by following the time evolution of an initial single $\Phi$ particle state. For this purpose we now introduce a method that implements a dynamical resummation directly in time\cite{dbzeno,infra} and is complementary to the dynamical renormalization group\cite{gold,boyvega}. We briefly revisit here the main aspects of this method for coherence and completeness of presentation, referring the reader to previous studies\cite{dbzeno,infra} for more details.

Consider a system whose Hamiltonian is $H=H_0+H_I$ with  $H_I$ a perturbation. The time evolution of states in the interaction picture
of $H_0$ is given by
\be i \frac{d}{dt}|\Psi_I(t)\rangle   = H_I(t)\,|\Psi_I(t)\rangle \,,  \label{intpic}\ee
where the interaction Hamiltonian in the interaction picture is
\be H_I(t) = e^{iH_0\,t} H_I e^{-iH_0\,t} \,. \label{HIoft}\ee

The Schroedinger eqn. (\ref{intpic}) has the formal solution
\be |\Psi_I(t)\rangle  = U(t,t_0) |\Psi_I(t_0)\rangle \,, \label{sol}\ee
   and the time evolution operator in the interaction picture $U(t,t_0)$ obeys \be i \frac{d}{dt}U(t,t_0)  = H_I(t)U(t,t_0)\,. \label{Ut}\ee

Now we can expand the time evolved state as \be |\Psi_I(t)\rangle  = \sum_n C_n(t) |n\rangle \,, \label{expan}\ee where $|n\rangle$ are eigenstates of the unperturbed Hamiltonian, $H_0 \ket{n} = E_n\,\ket{n}$, and form a complete set of orthonormal many particle states.   From eq.(\ref{intpic}) one finds the {\em exact} equation of motion for the coefficients $C_n(t)$, namely

\be \dot{C}_n(t) = -i \sum_m C_m(t) \langle n|H_I(t)|m\rangle \,. \label{eofm}\ee

Although this equation is exact, it generates an infinite hierarchy of simultaneous equations when the Hilbert space of states spanned by $\{|n\rangle\}$ is infinite dimensional. However, this hierarchy can be truncated by considering the transition between states connected by the interaction Hamiltonian at a given order in $H_I$.

\begin{figure}[ht!]
\begin{center}
\includegraphics[height=3in,width=3in,keepaspectratio=true]{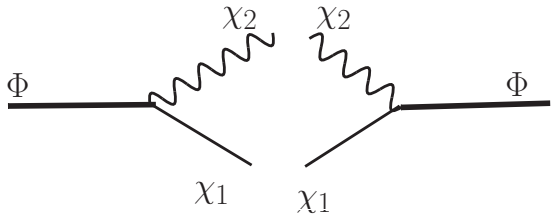}
\caption{Transitions $|1^{\Phi}_{\vk}\rangle \leftrightarrow |1^{\chi_1}_{\vp};1^{\chi_2}_{\vq}\rangle$ in first order in $H_I$.}
\label{fig1:coupling}
\end{center}
\end{figure}

Specifically, for the model under consideration here,
consider the situation depicted in figure (\ref{fig1:coupling}) where the single particle state, $|1^{\Phi}_{\vk}\rangle$, couples to the two particle state $|1^{\chi_1}_{\vp};1^{\chi_2}_{\vq}\rangle$, which couples back  to $|1^{\Phi}_{\vk}\rangle$ via the interaction Hamiltonian \be H_I(t) = \lambda\,\int d^3x\,\Phi(\vec{x},t)\,\chi_1(\vec{x},t)\,\chi_2(\vec{x},t)\,,\label{Hint}\ee where the fields are in the interaction picture.

Consider that at the initial time $t=0$ a single $\Phi$ particle state with momentum $\vk$ is prepared, upon time evolution the interaction Hamiltonian connects this state with a two particle state of the $\chi_1,\chi_2$ fields, therefore the time evolved state is given by

\be |\Psi_I(t)\rangle = C^{\Phi}_{\vk}(t)\,\ket{1^{\Phi}_{\vk}}+ \sum_{\vp}C^{\chi}_{\vp;\vk}(t)\,\ket{1^{\chi_1}_{\vp};1^{\chi_2}_{\vq}}+\cdots ~~;~~ \vec{q} = \vec{k}-\vec{p}\,, \label{psioft} \ee where the dots stand for multiparticle states that connect to $\ket{1^{\Phi}_{\vk}}$ in higher order in $H_I$, and we have explicitly used momentum conservation which is justified by the matrix elements obtained in appendix (\ref{app:sd}). In what follows we use $\vq = \vk-\vp$  to simplify notation.

The hierarchy of equations (\ref{eofm}) lead to the following coupled equations for the amplitudes
 \bea \dot{C}^{\Phi}_{\vk}(t) & = & -i \sum_{\vp} \langle 1^{\Phi}_{\vk}|H_I(t) |1^{\chi_1}_{\vp};1^{\chi_2}_{\vq}\rangle \,C^{\chi}_{\vp;\vk}(t)\label{Cfi}\\
\dot{C}^{\chi}_{\vp;\vk}(t) & = & -i \, C^{\Phi}_{\vk}(t) \langle 1^{\chi_1}_{\vp};1^{\chi_2}_{\vq}|H_I(t) |1^{\Phi}_{\vk}\rangle \,. \label{Cchis}\eea

The initial value problem in which at time $t=0$ the initial state is a single $\Phi$ particle state, namely $|\Psi(t=0)\rangle = | 1^{\Phi}_{\vk}\rangle $ and the vacuum for the other fields corresponds to the initial conditions
 \be C^{\Phi}_{\vk}(0)= 1,\   C^{\chi}_{\vp;\vk}(0) =0 .\label{initial}\ee  We   solve eq.(\ref{Cchis}) with these initial conditions  and input the solution into eq.(\ref{Cfi}) to find
\bea  C^{\chi}_{\vp;\vk}(t) & = &  -i \,\int_0^t \langle 1^{\chi_1}_{\vp};1^{\chi_2}_{\vq} |H_I(t')|1^{\Phi}_{\vk}\rangle \,C^{\Phi}_{\vk}(t')\,dt' \label{Ckapasol}\\ \dot{C}^{\Phi}_{\vk}(t) & = & - \int^t_0 \Sigma_{\Phi}(t,t') \, C^{\Phi}_{\vk}(t')\,dt' \label{intdiff} \eea where
\be \Sigma_{\Phi}(t,t') = \sum_{\vp} \langle 1^{\Phi}_{\vk}|H_I(t)|1^{\chi_1}_{\vp};1^{\chi_2}_{\vq}\rangle \langle 1^{\chi_1}_{\vp};1^{\chi_2}_{\vq}|H_I(t')|1^{\Phi}_{\vk}\rangle = \sum_{\vp} e^{i(E^{\Phi}_k- E^{\chi_1}_{\vp}-E^{\chi_2}_{\vq})(t-t')}\,|\bra{1^{\Phi}_{\vk}}H_I(0)\ket{1^{\chi_1}_{\vp};1^{\chi_2}_{\vq}}|^2\, \, \label{sigma} \ee and we used eqn. (\ref{HIoft}). It is convenient to write $\Sigma_{\Phi}(t,t') $ in a spectral representation, namely
\be \Sigma_{\Phi}(t,t') = \int^{\infty}_{-\infty}\,\rho_{\Phi}(k_0)\,e^{-i(k_0-E^{\Phi}_k)(t-t')}\,  dk_0 \label{specrep}\,,\ee where we have introduced the spectral density
\be \rho_{\Phi}(k_0) = \sum_{\vp}  \,|\bra{1^{\Phi}_{\vk}}H_I(0)\ket{1^{\chi_1}_{\vp};1^{\chi_2}_{\vq}}|^2\,\delta(k_0-E^{\chi_1}_{\vp}-E^{\chi_2}_{\vq})\,,  \label{rhopo}\ee which is obtained for the general case described by the Lagrangian density (\ref{lagsuper}) in appendix (\ref{app:sd}) (see eqn. (\ref{rhofina})).

The  integro-differential equation  with {\em memory} (\ref{intdiff}) yields a non-perturbative solution for the time evolution of the amplitudes and probabilities. It provides a resummation in real time of the one-particle irreducible self-energy corrections, akin to the
Dyson (geometric series) resummation of similar terms in the Fourier transform of the single particle propagator.

Inserting the solution for $C^{\Phi}_{\vk}(t)$ into eq.(\ref{Ckapasol}) one obtains the time evolution of amplitudes $C^{\chi}_{\vp;\vk}(t)$ from which we can compute  the time dependent probability to populate the two particle state $|1^{\chi_1}_{\vp};1^{\chi_2}_{\vq}\rangle$, namely $|C^{\chi}_{\vp;\vk}(t)|^2$.

The hermiticity of the interaction Hamiltonian $H_I$, and the equations (\ref{Cfi},\ref{Cchis}) yield
\be \frac{d}{dt} \Big[|C^{\Phi}_{\vk}(t)|^2 + \sum_{\vp}|C^{\chi}_{\vp;\vk}(t)|^2  \Big] = 0 \label{derit}\ee which  together with the initial conditions in eqs.(\ref{initial}) yields the unitarity relation
\be |C^{\Phi}_{\vk}(t)|^2 + \sum_{\vp}|C^{\chi}_{\vp;\vk}(t)|^2  =1\,. \label{unitarity1}\ee This  is the statement that the time evolution operator $U(t,0)$ is unitary, namely
\bea \langle \Psi_I(t)|\Psi_I(t)\rangle  & = & |C^{\Phi}_{\vk}(t)|^2 + \sum_{\vp}|C^{\chi}_{\vp;\vk}(t)|^2   \nonumber \\
 & =  & \langle \Psi(0)U^{\dagger}(t,0)   U(t,0) \Psi(0) \rangle = \langle \Psi(0)| \Psi(0) \rangle  =1 \,.  \label{unistat}\eea

The integro-differential  equation (\ref{intdiff}) can be solved exactly via Laplace transform\cite{dbzeno}, however finding the time evolution from the inverse transform involves a technically difficult integral with branch cut singularities. Instead, recognizing that for weak coupling there is a separation of time scales we invoke  the dynamical resummation method introduced in ref.\cite{infra} which   hinges on a separation of time scales warranted for weak coupling, and  provides a non-perturbative resummation directly in real time equivalent to the dynamical renormalization group\cite{gold,boyvega}.

The time evolution of $C^{\Phi}_{\vk}(t)$ determined by eq.(\ref{intdiff}) is \emph{slow} in the sense that
the time scale is determined by a weak coupling kernel $\Sigma$ which is second order in the coupling. This allows us to use an approximation in terms of a
consistent expansion in time derivatives of $C^{\Phi}_{\vk}(t)$. Let us define \be W_0(t,t') = \int^{t'}_0 \Sigma_{\Phi}(t,t'')dt'' \label{Wo}\ee so that \be \Sigma_{\Phi}(t,t') = \frac{d}{dt'}W_0(t,t'),\quad W_0(t,0)=0. \label{rela} \ee Integrating by parts in eq.(\ref{intdiff}) we obtain \be \int_0^t \Sigma_{\Phi}(t,t')\,C^{\Phi}_{\vk}(t')\, dt' = W_0(t,t)\,C^{\Phi}_{\vk}(t) - \int_0^t W_0(t,t')\, \frac{d}{dt'}C^{\Phi}_{\vk}(t') \,dt'. \label{marko1}\ee The second term on the right hand side is formally of \emph{fourth order} in $H_I$ suggesting how  a systematic approximation scheme can be developed. Setting \be W_1(t,t') = \int^{t'}_0 W_0(t,t'') dt'', \Rightarrow \frac{d}{dt'} W_1(t,t')= W_0(t,t');  \quad W_1(t,0) =0 \,\label{marko2} \ee and integrating by parts again, we find \be \int_0^t W_0(t,t')\, \frac{d}{dt'}C^{\Phi}_{\vk}(t') \,dt' = W_1(t,t)\,\dot{C}^{\Phi}_{\vk}(t) +\cdots \label{marko3} \ee leading to   \be \int_0^t \Sigma_{\Phi}(t,t')\,C^{\Phi}_{\vk}(t')\, dt' = W_0(t,t)\,C^{\Phi}_{\vk}(t) - W_1(t,t)\,\dot{C}^{\Phi}_{\vk}(t) +\cdots \label{histoint}\ee

This process can be implemented systematically resulting in higher order differential equations. Since $W_1 \simeq H^2_I $ and also $ \dot{C}_A \simeq H^2_I$ the second term in (\ref{histoint}) is $\simeq H^4_I$. We consistently neglect this term because to   order $H^4_I$ the states $\ket{1^{\chi_1}_{\vp};1^{\chi_2}_{\vq}}$   also have non-vanishing matrix elements with multiparticle states   other than $\ket{1^{\Phi}_{\vk}}$. These are the multiparticle states denoted by the dots in eqn. (\ref{psioft}) and the hierarchy would have to include these other states, therefore yielding contributions of $\mathcal{O}(H^4_I)$.  Hence up to   order $\simeq H^2_I$   the equation eq.(\ref{intdiff}) becomes \be \dot{C}^{\Phi}_{\vk}(t)= -  W_0(t,t) C^{\Phi}_{\vk}(t) \,,  \label{markovian}\ee   and from eqns. (\ref{specrep},\ref{Wo}) we find
\be W_0(t,t) = \int^{\infty}_{-\infty} \rho_{\Phi}(k_0) \Bigg[\frac{1-e^{-i(k_0-E^{\Phi}_k)t}}{i(p_0-E^{\Phi}_k)} \Bigg]\,dk_0\,,\label{Wzerot}\ee yielding
\be C^{\Phi}_{\vk}(t) = e^{-it\,\delta E^{\Phi}(t)  }\, e^{-\frac{\gamma(t)}{2}} \,, \label{solumarkov}\ee where we used the initial condition $C^{\Phi}_{\vk}(0)=1$,  with
\be \delta E^{\Phi}(t)  = \int^{\infty}_{-\infty} \frac{\rho_{\Phi}(k_0)}{(E^{\Phi}_k-k_0)}\,\Bigg[1- \frac{\sin\Big(\big(E^{\Phi}_k-k_0 \big)t\Big) }{(E^{\Phi}_k-k_0)t} \Bigg]\, dk_0\,, \label{real} \ee and
\be \gamma(t) = 2\,\int^{\infty}_{-\infty} \rho_{\Phi}(k_0) \frac{\Big[1-\cos\Big(\big(E^{\Phi}_k-k_0 \big)t\Big) \Big]}{\big(E^{\Phi}_k-k_0\big)^2}\,dk_0 \,. \label{imag} \ee
With this solution we find the time evolution of the coefficients of the multiparticle states from eqn. (\ref{Ckapasol})
\be C^{\chi}_{\vp;\vk}(t)   =    -i \,\langle 1^{\chi_1}_{\vp};1^{\chi_2}_{\vq} |H_I(0)|1^{\Phi}_{\vk}\rangle \,\int_0^t  \, e^{-i(E^{\Phi}_k- E^{\chi_1}_{\vp}-E^{\chi_2}_{\vq})t'}\, e^{-it'\,\delta E^{\Phi}(t')  }\, e^{-\frac{\gamma(t')}{2}}\,dt'\,, \label{soluchicoefs}\ee from which we obtain the probability of the multiparticle states in the time evolved state, and in particular the asymptotic state as $t\rightarrow \infty$.

 The survival probability of the initial state is given by
\be |\langle{1^{\Phi}_{\vk}}|{\Psi(t)}\rangle|^2 = |C^{\Phi}_{\vk}(t)|^2 = e^{-\gamma(t)}\,. \label{surviprob}\ee

In the long time limit we find

\be \delta E^{\Phi}(t)_{~~ \overrightarrow{t\longrightarrow \infty}~~} \,\delta E^{\Phi}_\infty = \int^{\infty}_{-\infty} \mathcal{P}\, \frac{\rho_{\Phi}(k_0)}{(E^{\Phi}_{\vk}-k_0)}\,dk_0 \,,\label{asyreal}\ee  where $\mathcal{P}$ stands for the principal part,  yielding a renormalization of the bare frequency of the state $|1^{\Phi}_{\vk}\rangle$, namely $E^{\Phi}_{\vk}+ \delta E^{\Phi}_\infty = E^{\Phi}_{\vk;R}$ , whereas the long time limit of $\gamma(t)$   yields the decay law of the initial state.

It is illuminating to write   the energy renormalization using the explicit form of the spectral density given by eqn. (\ref{rhopo}), namely
\be \delta E^{\Phi}_\infty = {\sum_{\vp}}' \frac{|\bra{1^{\Phi}_{\vk}}H_I(0)\ket{1^{\chi_1}_{\vp};1^{\chi_2}_{\vq}}|^2 }{E^{\Phi}_{\vk}- E^{\chi_1}_{\vp}-E^{\chi_2}_{\vq}} \,,\label{renergy}\ee where the principal part in eqn.(\ref{asyreal}) removes the region in momenta when the denominator vanishes denoted by the superscript prime in the sum. This is the usual quantum mechanical result for the second order energy shift.

\vspace{1mm}

\subsection{Stable particles:}\label{subsec:stables}

Before we analyze the time evolution of the coefficients  $C^{\Phi}_{\vk}(t)$, we can understand their asymptotic behavior for the case of stable particles.
The spectral density $\rho_{\Phi}(k_0)$   vanishes for $k_0 < k_{0T}$ where $k_{0T}=\sqrt{k^2+(m_1+m_2)^2}$   corresponds to the two particle threshold (see eqn. (\ref{rhofina})).
In the case of a stable particle $E^{\Phi}_k < k_{0T}$ the denominator in $\gamma(t)$, eqn. (\ref{imag}) never vanishes and the cosine term averages out in the long time limit. Therefore in the case of a stable particle with energy below the two particle threshold energy it follows that
\be \gamma(t)_{~~ \overrightarrow{t\longrightarrow \infty}~~} 2\,\int^{\infty}_{k_{0T}}   \frac{\rho_{\Phi}(k_0) }{\big(E^{\Phi}_k-k_0\big)^2}\,dk_0  \equiv 2 z \,. \label{lilz}\ee Hence,  in the case of a stable particle for which the single particle energy is below the two particle threshold (neglecting renormalization to lowest order)   the time evolution of the initial single $\Phi$ particle amplitude yields the asymptotic result
\be C^{\Phi}_{\vk}(\infty) = e^{-it\,\delta E^{\Phi}_\infty  }\, e^{-z} \,, \label{asyCfi} \ee namely, the probability of finding the initial (bare) single particle state in the asymptotic state $|\Psi_I(\infty)\rangle$ is
\be |\langle 1^{\Phi}_{\vk}|\Psi_I(\infty)\rangle|^2 = |C^{\Phi}_{\vk}(\infty)|^2   = e^{-2z}\equiv \mathcal{Z} \,,  \label{zamp}\ee and the unitarity condition (\ref{unitarity1}) implies the sum rule
\be \mathcal{Z} + \sum_{\vp}|C^{\chi}_{\vp;\vk}(\infty)|^2  = 1\,, \label{sumrul}\ee we show how  this relation is fulfilled in section (\ref{sec:uniasy}) (see eqn. (\ref{stableproof})).

For the case $m_1=m_2=m$, it is straightforward to find that as $M^2\rightarrow 4 m^2$, namely as $E^{\Phi}_k \rightarrow k_{0T}$ the integral in (\ref{lilz}) yields $z \propto (4m^2-M^2)^{-1/2}$ displaying the threshold divergence that results in the vanishing of the overlap between the asymptotic state and the initial single particle state, namely $\mathcal{Z}\rightarrow 0$.

As we discuss below this is a consequence of taking the infinite time limit too soon in the limit when the single particle energy approaches the two particle threshold.

It is illuminating to understand the time scale over which the integral (\ref{imag}) approaches its asymptotic limit (\ref{lilz}). In the case of a stable particle, with $E_k < k_{0T}$ as mentioned above, the denominator in (\ref{imag}) does not vanish in the domain of integration $k_0 \geq k_{0T}$, therefore we can separate the time dependent cosine term from the expression for $\gamma(t)$. Hence,  consider the integral
\be \mathcal{I}(t) = \int^{\infty}_{k_{0T}} \rho_{\Phi}(k_0) \frac{  \cos\Big(\big(E^{\Phi}_k-k_0 \big)t\Big) }{\big(E^{\Phi}_k-k_0\big)^2}\,dk_0 \,, \ee in the long time limit the cosine  averages out and this integral vanishes by dephasing, on a time scale
\be t_{dp} \simeq \frac{1}{k_{0T}-E^\Phi_k}\ee since $k_{0T}- E_k$ is the smallest frequency contributing to the integral, which, in turn dominates the long time limit. Therefore as the single particle energy approaches the threshold from below the dephasing time scale $t_{dp}$ diverges, and as discussed above the overlap $\mathcal{Z}$ vanishes.
This is the case of threshold divergences, the dynamics of which will be studied in detail below.

\vspace{1mm}

\subsection{Decaying particle:}\label{subsec:decayparts}

 In this case the (renormalized) single particle energy is above the two particle threshold, $E^{\Phi}_k > k_{0T}$, and the denominator in (\ref{imag}) vanishes within the domain of integration, therefore the cosine term cannot be separated.
Let us consider the case of equal masses $m_1=m_2=m$ in which case we find from eqns. (\ref{imag}, \ref{rhofina}) that
\be \gamma(t) = g^2  \,\frac{M^2}{E^{\Phi}_k}\,\int^{\infty}_{k_{0T}}\Bigg[ 1-\frac{4m^2}{k^2_0-k^2} \Bigg]^{1/2}\,  \frac{\Big[1-\cos\Big(\big(E^{\Phi}_k-k_0 \big)t\Big) \Big]}{\big(E^{\Phi}_k-k_0\big)^2}\,dk_0 ~~;~~ k_{0T} = \sqrt{k^2+4m^2}\,. \label{gamadecy} \ee
Define
\be (k_0-E^{\Phi}_k)t \equiv x \label{defx}\ee in terms of which
\be \gamma(t) = \frac{g^2\,M^2}{E^{\Phi}_k}\,  t\, \, \int^{\infty}_{-X(t)} \overline{\rho}(x/Mt)\,\frac{1-\cos(x)}{x^2} dx ~~;~~ X(t) =  (E^{\Phi}_k-k_{0T})\,t >0 \label{gamares}\ee where for a decaying state with $M^2> 4m^2  \Rightarrow E^{\Phi}_k > k_{0T}$, and
\be \overline{\rho}(\xi)=   \Bigg[ \frac{1-\frac{4m^2}{M^2} +  \frac{2E^{\Phi}_k}{M}\,\xi+ \xi^2  }{  1+\frac{2E^{\Phi}_k}{M}\,\xi+ \xi^2 } \Bigg]^{1/2}~~;~~ \xi = \frac{x}{Mt}
 \,. \label{overho}\ee   In the long time limit $Mt\rightarrow \infty;X(t)\rightarrow \infty$  we find
\be \gamma(t) \rightarrow   \, \Gamma_k \, t +2\,z_d + \mathcal{O}(1/t) \,,\label{decayte}\ee where
 \be  \Gamma_k  = \pi g^2 M \,\Bigg[1- \frac{4m^2}{M^2 } \Bigg]^{1/2}\,\Big(\frac{M}{E^\Phi_k}\Big)= 2\pi \,\rho_\Phi(E^\Phi_k) \,, \label{decrat}\ee  is the correct decay rate (\ref{gammapole}) including the time dilation factor, and is identified with the usual result from Fermi's golden rule,  and
\be 2\,z_d = \frac{ g^2 M }{E^{\Phi}_k}\,\Bigg\{-\frac{2\,\overline{\rho}(0)}{  \overline{\xi}}\,+  \, \int^{\overline{\xi}}_{-\overline{\xi}} \Big[\overline{\rho}_e(\xi)- \overline{\rho}(0) \Big]\,    \frac{d\xi}{\xi^2} + \int^{\infty}_{\overline{\xi}} \frac{\overline{\rho}(\xi)}{\xi^2}\,    {d\xi}  \Bigg\}~~;~~ \overline{\xi} = (E^{\Phi}_k-k_{0T})/M \,,  \ee where $\overline{\rho}_e(\xi) = ( \overline{\rho}(\xi)+ \overline{\rho}(-\xi))/2$. The details of the derivation of this result are given in appendix (\ref{app:longama}).
We find     the long time behavior
\be  C^{\Phi}_{\vk}(t) ~~~~{}_{\overrightarrow{t\rightarrow \infty}} ~~e^{-i\delta E^{\Phi}_\infty\,t }\,e^{-\frac{\Gamma_k}{2}\, t} \,e^{-z_d} \Rightarrow |C^{\Phi}_{\vk}(t)|^2~~~~{}_{\overrightarrow{t\rightarrow \infty}} ~~\mathcal{Z}_d\,e^{-\Gamma_k t}\,,~~;~~ \mathcal{Z}_d=e^{-2z_d}  \label{deca}\ee
where $\delta E^{\Phi}_\infty$ is a renormalization of the   single particle energy.

\vspace{1mm}

\subsection{Threshold singularity:}\label{subsec:threshold}

The   expression for $\gamma(t)$, eqn. (\ref{gamares}) in terms of $\overline{\rho}$ given by eqn. (\ref{overho}) makes explicit the modification of the decay   in the case of threshold singularity, namely $M^2=4m^2$, in this case,
\be \overline{\rho}(\xi)=   \Bigg[ \frac{ \frac{2E^{\Phi}_k}{M}\,\xi+ \xi^2  }{  1+\frac{2E^{\Phi}_k}{M}\,\xi+ \xi^2 } \Bigg]^{1/2}
 ~~;~~ \xi =   \frac{x}{Mt} \,, \label{overhoth}\ee we note that in this case the decay rate (\ref{decrat}) $\Gamma_k=0$, and the spectral density vanishes with a square root at threshold.

 In the limit $t \rightarrow \infty$ it follows that
$\overline{\rho} \rightarrow \big[ \frac{2E^{\Phi}_k\,x}{M^2t}\big]^{1/2}$ yielding
\be \gamma(t) = 2\,\sqrt{\pi}\,g^2 \,\sqrt{\frac{M}{E^\Phi_k}} \,\sqrt{Mt}\,,  \label{gamath}\ee  and the survival probability decays as
\be |C^{\Phi}_{\vk}(t)|^2~~~~{}_{\overrightarrow{t\rightarrow \infty}} ~~e^{-\sqrt{t/t^*_k}}\,, \label{thresdec}\ee
with an effective lifetime
\be t^*_k = \frac{1}{4\pi\,g^4\,M}\,\frac{E^\Phi_k}{M}\,, \label{lifethresh}\ee where   $E^\Phi_k/M$ is the usual time dilation factor.
Namely the decay law changes from $e^{-\Gamma_k \,t} \rightarrow e^{-  \sqrt{t/t^*_k}}$,  as the mass of the particle approaches the threshold. The square root behavior is a consequence of fact that the spectral density vanishes as a square root near threshold.

Furthermore, whereas in the case of decay there is a constant, time independent contribution in the asymptotic long time limit of $\gamma(t)$ which defines the  wave function renormalization, no such term arises in the case of threshold singularity.
 Therefore, at threshold when $M^2=4m^2$, even when the usual decay rate (\ref{gammapole}) vanishes, the amplitude  of the initial single particle state decays  not as $e^{-\Gamma t}$  but as $e^{-\sqrt{t/t^*}}$ with an effective lifetime $t^*$ given by eqn. (\ref{lifethresh}), reflecting the square root divergence in both in the spectral density approaching threshold from above and the wave function renormalization approaching the threshold from below. This new decay law is in qualitative agreement with a result found in ref.\cite{thres} and  implies that the single particle state is not an asymptotic state in agreement with a vanishing wave function renormalization from below, and the fact that the continuum contribution of the spectral density saturates the sum rule.

In order to understand the asymptotic behavior in more detail it proves convenient to study the case $k=0$ and to introduce the
dimensionless combinations $r= 4m^2/M^2 $ and $\tau = M\,t$, in terms of which, for $k=0$,
\be \gamma(t) = g^2\,\tau \,J(r,\tau) \label{kzero} \ee where
\be J(r,\tau) = \int^{\infty}_{-(1-r)\,\tau} \Bigg[\frac{(1+\frac{x}{\tau})^2-r}{(1+\frac{x}{\tau})^2} \Bigg]^{1/2}\,\frac{1-\cos(x)}{x^2}\,,dx \,.
\label{jfunc}\ee

\begin{figure}[h!]
\begin{center}
\includegraphics[height=4in,width=4in,keepaspectratio=true]{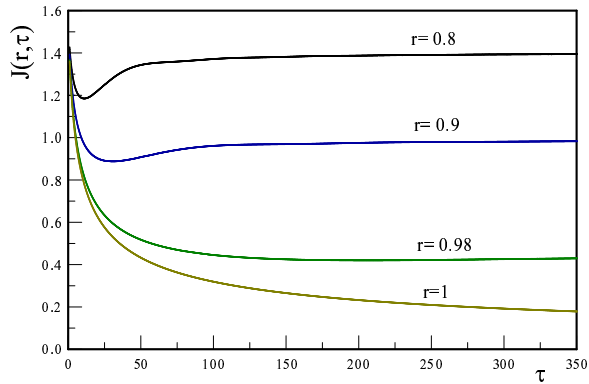}
\caption{The function $J(r,\tau)$ with $r=4m^2/M^2~~;~~ \tau=M\,t$ for $r, 0.8,0.9,0.98,1$.}
\label{fig9:asintote}
\end{center}
\end{figure}

Since the factor $(1-\cos(x))/x^2$ is localized within a region of width $\simeq 2\pi$ around the origin, for $r<1$ the function $J(r,\tau)$ approaches its asymptotic limit $J(r,\infty)= \pi\,(1-r)^{1/2}$ within a $\tau$ scale $\simeq 2\pi/(1-r)$. As the threshold is approached from above, namely $r\rightarrow 1$ from below, the asymptotic value becomes smaller and smaller taking a longer and longer time scale to reach, and for $r=1$ the function $J(1,\tau) \propto 1/\sqrt{\tau}$ for large $\tau$. This behavior is clearly displayed in fig. (\ref{fig9:asintote}) for $r=0.8,0.9,0.98,1$.

The crossover between the linear and square root behavior can be understood quantitatively in the intermediate asymptotic regime for
$(1-r) \ll 1$ from the following argument. Consider first the case $k=0$, and focus  on the numerator of the term within brackets in $J(r,\tau)$, eqn. (\ref{jfunc}). In the region  $(1-r)  \ll (2\pi/\tau) \ll  1$ the contribution  $2x/\tau$ dominates in the numerator yielding $J(r,\tau) \propto 1/\sqrt{\tau}$. This behavior continues until $(1-r) \gtrsim 2\pi/\tau$ at which point there is a crossover and the function $J(r,\tau)$ reaches the constant value $\pi \,(1-r)$. As $r\rightarrow 1$ from below, this constant value vanishes on a very long time scale $\propto 1/(1-r)$ during which $J(r,\tau)$ falls off $\propto 1/\sqrt{\tau}$. Therefore for $r < 1$ the crossover from the square root fall off to the asymptotic constant value occurs at a time scale $t_{x} \simeq 2\pi/M(1-r)$. For $k\neq 0$ this time scale is modified by the time dilation factor $E^\Phi_k/M$. In section (\ref{sec:discussion}) we comment on the effect of radiative corrections on threshold behavior.

\subsection{Infrared singularity:}\label{subsec:infra}

The case of infrared singularity corresponds to $m_1=M,m_2=0$,
where the spectral density (\ref{rhofina}) simplifies to
\be  \rho_{\Phi}(k_0) = g^2  \,\frac{M^2}{2E^{\Phi}_k}\,\Bigg[ \frac{(k_0-E^\Phi_k)(k_0+E^\Phi_k) }{k^2_0-k^2} \Bigg] \,\Theta(k_0- E^\Phi_k)\,, \label{IRrho} \ee vanishing linearly as $k_0$ approaches the threshold $k_{0T} = E^\Phi_k$. This situation must be contrasted with the case of threshold divergence where the spectral density vanishes as a square root as $k_0 $ approaches threshold. However, in both cases the usual decay rate $\Gamma$ given by eqn. (\ref{decrat}) vanishes.

It is convenient to introduce the dimensionless combinations
\be \eta =  \frac{k_0-E^\Phi_k}{E^\Phi_k}~~;~~\mathcal{T}= E^\Phi_k\,t \,, \label{eta}\ee in terms of which we find in this case
\be \gamma(t) = \Big(\frac{g \,M }{E^\Phi_k}\Big)^2\, \int^\infty_0 \frac{2+\eta}{\big(\frac{ M }{E^\Phi_k} \big)^2 +2\eta+\eta^2}\, \,\frac{1-\cos(\eta\,\mathcal{T})}{\eta} \, \,  d\eta \,.\label{gamaIR} \ee We note that this integral features a logarithimic divergence in the region of small $\eta$. Following ref. \cite{infra} we write the above integral as
\be \gamma(\mathcal{T}) = I_1(\mathcal{T})+I_2(\mathcal{T})\,, \label{intsplit}\ee with
\be I_1(\mathcal{T}) =  2 g^2 \, \int^1_0 \frac{1-\cos(\eta\,\mathcal{T})}{\eta} \, \,  d\eta  + g^2 \,\int^1_0  \Bigg[\frac{\big(\frac{ M }{E^\Phi_k} \big)^2-4-2\eta}{\big(\frac{ M }{E^\Phi_k} \big)^2+2\eta+\eta^2} \Bigg]\,(1-\cos(\eta\,\mathcal{T})) \, d\eta \,,\label{I1}\ee and
\be I_2(\mathcal{T}) = \Big(\frac{g \,M }{E^\Phi_k}\Big)^2\, \int^\infty_1 \frac{2+\eta}{\big(\frac{ M }{E^\Phi_k} \big)^2 +2\eta+\eta^2}\, \,\frac{1-\cos(\eta\,\mathcal{T})}{\eta} \, \,  d\eta \,.\label{I2} \ee In the long time limit $\mathcal{T} \gg 1$ the first integral in $I_1(\mathcal{T})$ features an infrared logarithmic divergence, whereas in the second integral and in $I_2(\mathcal{T})$ the cosine terms average out yielding in this limit
\be \gamma(t)~~{}_{\overrightarrow{\mathcal{T}\rightarrow \infty}} ~~ 2 g^2 \ln[ {E^\Phi_k\,t}] + 2 z_{ir}\,,\label{asyIR} \ee where
\be  z_{ir} = \frac{g^2 }{2}\Bigg\{ 2\gamma_E + \int^1_0  \Bigg[\frac{\big(\frac{ M }{E^\Phi_k} \big)^2-4-2\eta}{\big(\frac{ M }{E^\Phi_k} \big)^2+2\eta+\eta^2} \Bigg]  \, d\eta +  \Big(\frac{M }{E^\Phi_k}\Big)^2\, \int^\infty_1 \frac{2+\eta}{\big(\frac{ M }{E^\Phi_k} \big)^2 +2\eta+\eta^2}\,   \,  \frac{d\eta }{\eta}     \Bigg\}\,, \ee and $\gamma_E$ is the Euler-Mascheroni constant, $z_{ir}$ is infrared and ultraviolet finite. Therefore, for the infrared case we find
\be  |C^{\Phi}_{\vk}(t)|^2~~~~{}_{\overrightarrow{E^\Phi_k \,t \gg 1}} ~~\big[E^\Phi_k \,t\big]^{-2g^2}\,\mathcal{Z}_{ir} ~~;~~ \mathcal{Z}_{ir} = e^{-2z_{ir}} \,. \label{IRcoef}  \ee namely the probability of the initial single particle state decays in time as a power law with anomalous dimension $2g^2$ and is not an asymptotic state in   S-matrix amplitudes.

In summary, we find the following asymptotic long time limits for the unstable cases in which the ``mass shell'' of the particle is above or at threshold,

\be |C^{\Phi}_{\vk}(t)|^2~~~~{}_{\overrightarrow{t\rightarrow \infty}} ~~   \left\{ \begin{array}{l}
                                                                                 \mathcal{Z}_d\, e^{-\Gamma_k t}~~\textrm{ above threshold}  \\
                                                                                  e^{-\sqrt{t/t^*_k}} ~~\textrm{ at threshold~~;~~$m_2\neq 0$}   \\
                                                                                \mathcal{Z}_{ir}\big[E^\Phi_k \,t\big]^{-2g^2} ~~\textrm{ infrared~;~~$m_2 = 0$}
                                                                                \end{array}   \right.
\,  \label{allcases} \ee

\section{Unitarity and asymptotic state}\label{sec:uniasy}

\subsection{Unitarity:}\label{sub:unitarity}
In all cases of ``generalized decay'' as described by eqn. (\ref{allcases}) the asymptotic state is
\be |\Psi_I(\infty)\rangle =   \sum_{\vp}C^{\chi}_{\vp;\vk}(\infty)\,\ket{1^{\chi_1}_{\vp};1^{\chi_2}_{\vq}} \,, \label{asystatefina} \ee
 The probabilities must obey the sum rule
\be \sum_{\vp}|C^{\chi}_{\vp;\vk}(\infty)|^2 =1\,,\label{asysr} \ee which is the statement of unitarity  (\ref{unitarity1}) in the asymptotic long time limit when the amplitude of the initial state vanishes.

The question that we address is how   this sum rule is fulfilled being that
the probabilities $|C^{\chi}_{\vp;\vk}(\infty)|^2$ are formally of order $g^2$. In appendix (\ref{app:identity}) we show that for all cases, and up to $\mathcal{O}(H^4_I)$ the asymptotic probabilities are given by
\be |C^{\chi}_{\vp;\vk}(\infty)|^2 = \frac{2}{\Omega}\,|\langle 1^{\chi_1}_{\vp};1^{\chi_2}_{\vq} |H_I(0)|1^{\Phi}_{\vk}\rangle|^2 \, \int^{\infty}_0 \sin\big[\Omega\,t \big]\,e^{-\gamma(t )}\, dt  ~~;~~ \Omega \equiv  E^{\chi_1}_{\vp}+E^{\chi_2}_{\vq} - E^{\Phi}_k \,, \label{finprobchi2asy}\ee where $E^{\Phi}_k$ in this expression is the renormalized single particle energy (see eqn. (\ref{finprobchi2})). Introducing the spectral representation (\ref{rhopo}) we finally find the general form  valid up to $\mathcal{O}(H^4_I)$
\be \sum_{\vp}|C^{\chi}_{\vp;\vk}(\infty)|^2 = 2 \,\int^{\infty}_0  \int^{\infty}_{-\infty}\frac{\rho_{\Phi}(k_0)}{(k_0 - E^{\Phi}_k)}\, \sin\big[(k_0 - E^{\Phi}_k)\,t \big]\,e^{-\gamma(t )} \, dt \,  dk_0 \,. \label{general}\ee

Armed with this general expression we can now study the individual cases by considering the different forms of $\gamma(t)$ and spectral densities.

\vspace{1mm}

\textbf{Decay:}
 For the case of decay neglecting early time transient dynamics before the linear secular growth in time in the exponent, which only yields  a perturbative contribution, $\gamma(t)$ is given by (\ref{decayte}) and $\rho_{\Phi}(k_0) $ by eqn. (\ref{rhofina}) for the case $m_1=m_2 = m$. In this case the time integral is straightforward leading to
 \be \sum_{\vp}|C^{\chi}_{\vp;\vk}(\infty)|^2 = 2\, \mathcal{Z}_d\,  \int^{\infty}_{-\infty}\frac{\rho_{\Phi}(k_0)}{\big[k_0 - E^{\Phi}_k\big]^2+ \Gamma^2_k}\,d k_0 \,. \label{c2infidecay} \ee
 We can confirm the unitarity relation (\ref{asysr}) to leading (zeroth) order at this stage by setting $\mathcal{Z}_d=1$ and writing
 \be \sum_{\vp}|C^{\chi}_{\vp;\vk}(\infty)|^2 = \frac{2\,\pi}{\Gamma_k} \,    \int^{\infty}_{-\infty}\rho_{\Phi}(k_0)\,\frac{1}{\pi}\frac{\Gamma_k}{\big[k_0 - E^{\Phi}_k\big]^2+ \Gamma^2_k}\,d k_0 \,. \label{unisimple} \ee In the narrow width limit
 \be \frac{1}{\pi}\frac{\Gamma_k}{\big[k_0 - E^{\Phi}_k\big]^2+ \Gamma^2_k} ~~ {}_{\overrightarrow{\Gamma_k \rightarrow 0}}~~\delta(k_0 - E^{\Phi}_k)\,, \label{delfunap}\ee yielding
 \be \sum_{\vp}|C^{\chi}_{\vp;\vk}(\infty)|^2 = \frac{2\,\pi}{\Gamma_k} \, \rho_{\Phi}(E^{\Phi}_k) =1\,, \label{narwid}\ee where we used the result (\ref{decrat}). To prove unitarity up to $\mathcal{O}(H^4_I)$ requires a somewhat deeper analysis, which we now undertake.

  Let us introduce the dimensionless variables
\be \xi = \frac{k_0 - E^{\Phi}_k}{M}~~;~~ \varepsilon = \frac{\Gamma_k}{M} ~~;~~ \overline{\xi} = \frac{E^{\Phi}_k - k_{0T}}{M}>0 \,, \label{varsdi}\ee it follows that
\be \sum_{\vp}|C^{\chi}_{\vp;\vk}(\infty)|^2 = \mathcal{Z}_d\, \frac{g^2\,M}{E^{\Phi}_k }\, \int^{\infty}_{-\overline{\xi}} \frac{\overline{\rho}(\xi)}{\xi^2+\varepsilon^2}\, d\xi \,,\label{decform} \ee with $\overline{\rho}(\xi)$ given by eqn. (\ref{overho}).

Following similar steps as in appendix (\ref{app:longama}) we find
\be \sum_{\vp}|C^{\chi}_{\vp;\vk}(\infty)|^2 =   \mathcal{Z}_d\, \frac{g^2\,M}{E^{\Phi}_k }\, \Bigg\{ \overline{\rho}(0)\, \int^{\overline{\xi}}_{-\overline{\xi}} \frac{1}{\xi^2+\varepsilon^2}\, d\xi + \int^{\overline{\xi}}_{-\overline{\xi}} \frac{\big[ \overline{\rho}_e(\xi)-\overline{\rho}(0) \big]}{\xi^2+\varepsilon^2}\, d\xi + \int^{\infty}_{ \overline{\xi}} \frac{ \overline{\rho}(\xi)}{\xi^2+\varepsilon^2}\,   d\xi   \Bigg\}\,. \label{expres}\ee In the narrow width limit $\varepsilon \ll 1$ the first integral is straightforward yielding $\pi/\varepsilon - 2/ \overline{\xi} + \mathcal{O}(\varepsilon)$,   and for $ \overline{\xi}\gg \varepsilon$  ($ E^\Phi_k -k_{0T} \gg \Gamma_k$) we can set $\varepsilon \rightarrow 0$ in the second and   third integrals\footnote{The second integral is finite in this limit because $\overline{\rho}_e(\xi)-\overline{\rho}(0) \simeq \xi^2$ as $\xi \rightarrow 0$.},   yielding
\be \sum_{\vp}|C^{\chi}_{\vp;\vk}(\infty)|^2 =   \mathcal{Z}_d\,\big[1+2z_d \big]\,,\label{unideca1}\ee where we used the results (\ref{gamapp},\ref{zdapp}). Therefore, with $\mathcal{Z}_d = e^{-2z_d} \simeq 1-2z_d + \cdots$ we indeed find that the unitarity relation (\ref{asysr}) is fulfilled up to $\mathcal{O}(H^4_I)$ consistently with our main approximation.

From this result we can confirm unitarity also in the stable case, namely eqn. (\ref{sumrul}), simply by taking $\mathcal{Z}_d \rightarrow \mathcal{Z}$, the amplitude of the single particle contribution in eqn. (\ref{zamp}), and  the limit $\Gamma \rightarrow 0$ which is non-singular in eqn. (\ref{c2infidecay}) because $E^{\Phi}_k< k_{0T}$, therefore for $\Gamma =0$ the denominator in eqn. (\ref{c2infidecay}) never vanishes, furthermore, only the last term  inside the brackets in eqn. (\ref{expres}) contributes in the stable case because $E^{\Phi}_k< k_{0T}$.  $\Gamma \rightarrow 0^+$ fulfills the role of a convergence factor in the integral in (\ref{finprobchi2asy}). Including the contribution from the single particle state with weight $\mathcal{Z}$ we find
\be \mathcal{Z}\Big[1+ 2 z \Big] =1 \,,\label{stableproof} \ee where $2z$ is given by eqn. (\ref{lilz}), thus proving the sum rule (\ref{sumrul}) up to $\mathcal{O}(H^4_I)$ in the stable case.

\vspace{1mm}

\textbf{Threshold singularity:} In this case $\gamma(t)$ is given by eqn. (\ref{gamath}), and $\rho_\Phi(k_0)$ is given by eqn. (\ref{rhofina}) with $m_1=m_2=m~~;~ 4m^2 = M^2$. In the general expression (\ref{general}) we introduce the following variables
\be (k_0-E^\Phi_k) = E^\Phi_k \, s~~;~~  t = \frac{\pi\,u}{2\,E^\Phi_k \, s}  ~~;~~ \delta = \frac{\pi\,g^2 M}{E^\Phi_k}~~;~~\beta(s) = \delta\,\frac{\sqrt{2}}{\sqrt{s}} \,,\label{thvari}\ee
obtaining
\be \sum_{\vp}|C^{\chi}_{\vp;\vk}(\infty)|^2 = \delta\, \int^{\infty}_{0} \Bigg[\frac{2+s}{1+ 2\,\big(\frac{E^\Phi_k}{M}\big)^2 \,s + \big(\frac{E^\Phi_k}{M}\big)^2 \,s^2} \Bigg]^{1/2}\, \frac{ds}{s^{3/2}}\,\int^\infty_0 \sin\Big[\frac{\pi\,u^2}{2} \Big]\,e^{-\beta(s)\,u}\,u\,du \,.  \ee Finally, we rescale the coupling by writing
\be s \equiv \delta^2 \, y \,, \label{rescath}\ee yielding
\be \sum_{\vp}|C^{\chi}_{\vp;\vk}(\infty)|^2 =   \int^{\infty}_{0} \Bigg[\frac{2+\delta^2 \,y}{1+ 2\,\delta^2\,\big(\frac{E^\Phi_k}{M}\big)^2 \,y + \delta^4\, \big(\frac{E^\Phi_k}{M}\big)^2 \,y^2} \Bigg]^{1/2}\, \frac{dy}{y^{3/2}}\,\int^\infty_0 \sin\Big[\frac{\pi\,u^2}{2} \Big]\,e^{-\sqrt{\frac{2}{y}}\,\,u}\,u\,du \,. \label{C2ir}  \ee Because $\delta^2 \propto g^4 \propto H^4_I$ up to this order we can set $\delta=0$ in the above expression, the remaining integrals are elementary\footnote{By a change of variables $y^{-1/2} \equiv x$, the resulting integral yields the Sine Fresnel integral. } yielding
\be \sum_{\vp}|C^{\chi}_{\vp;\vk}(\infty)|^2   =1\,, \ee thus confirming the unitarity relation  (\ref{asysr}) up to $\mathcal{O}(H^4_I)$ in this case.

\vspace{1mm}

\textbf{Infrared divergence:}
The fulfillment of the unitarity condition (\ref{asysr}) in the case of infrared divergence has been confirmed up to $\mathcal{O}(H^4_I)$ in ref.\cite{infra} to which the reader is referred   for further technical details. However, for the sake of completeness we here summarize the main steps to leading (zeroth) order to compare with the previous cases. In this case, the spectral density is given by eqn. (\ref{rhofina}) in appendix (\ref{app:sd}) with $m_1=M, m_2=0$, which when combined with the general result (\ref{general}) and the result (\ref{asyIR}) for $\gamma(t)$ and setting $z_{ir}=0$ to leading order, yields
\be \sum_{\vp}|C^{\chi}_{\vp;\vk}(\infty)|^2 = 2 \int \frac{\rho_\Phi(k_0)}{\big[k_0 - E^\Phi_k \big]^2}\,\Bigg[\frac{k_0 - E^\Phi_k }{E^\Phi_k} \Bigg]^{2g^2}\,dk_0 ~~ \underbrace{\int^\infty_0 \sin[\tau]\,\tau^{-2g^2} \,d\tau }_{=1+ \mathcal{O}(g^2)}\,.  \label{IRc2s}\ee
Changing variables to $s= (k_0 - E^\Phi_k)/E^\Phi_k$ we find to leading order
\be   \sum_{\vp}|C^{\chi}_{\vp;\vk}(\infty)|^2 =   g^2 \,\Big(\frac{M}{E^\Phi_k}\Big)^2\, \int^{\infty}_0 \Bigg[\frac{2+s}{\Big(\frac{M}{E^\Phi_k}\Big)^2+2s+s^2}\Bigg]\,s^{2g^2} \,\frac{ds}{s}\,, \label{nuc2ir} \ee writing $\int^\infty_0 (\cdots) ds = \int^1_0 (\cdots) ds + \int^\infty_1 (\cdots) ds $ and in the first integral  separating the infrared dominant term by writing
\be \frac{2+s}{\Big(\frac{M}{E^\Phi_k}\Big)^2+2s+s^2} = \frac{2}{\Big(\frac{M}{E^\Phi_k}\Big)^2}+ \frac{s}{\Big(\frac{M}{E^\Phi_k}\Big)^2}\,\Bigg[ \frac{\Big(\frac{M}{E^\Phi_k}\Big)^2 -4-2s}{\Big(\frac{M}{E^\Phi_k}\Big)^2+2s+s^2}\Bigg] \,, \label{bigfor}\ee the second term in (\ref{bigfor}) above along with the integral $\int^\infty_1 (\cdots) ds $ yield contributions of order $\mathcal{O}(g^2)$,  the leading order term  is given by
\be \sum_{\vp}|C^{\chi}_{\vp;\vk}(\infty)|^2 = 2g^2 \,\int^1_0 s^{2g^2} \,\frac{ds}{s} =1 \label{unixc2ir} \ee confirming the unitarity constraint up to leading order, the details of the confirmation up to $\mathcal{O}(g^4)$ are available in ref.\cite{infra}.

\subsection{The asymptotic state:}\label{sub:asystate}
In  the  cases of decay, threshold and infrared singularities discussed above, the asymptotic state after the amplitude of the initial state has become negligible, features a common form, namely
\be |\Psi_I(\infty)\rangle =  \sum_{\vp}C^{\chi}_{\vp;\vk}(\infty)\,\ket{1^{\chi_1}_{\vp};1^{\chi_2}_{\vk-\vp}} \,,\label{comonasy}\ee
or in the case of the infrared singularity for the model given by the Lagrangian density (\ref{charged}) of a charged scalar field interacting with a massless scalar, obtained by identifying $\chi_1 \equiv \Phi~;~\chi_2\equiv \chi$ and $\chi$ a massless field, namely
\be |\Psi_I(\infty)\rangle =  \sum_{\vp}C^{\Phi\chi}_{\vp;\vk}(\infty)\,\ket{1^{\Phi }_{\vp};1^{\chi }_{\vk-\vp}} \,.\label{comonasyfichi}\ee In the analysis below we will consider the asymptotic state (\ref{comonasy}) describing all cases with the implicit understanding that the case of infrared divergence is obtained by the replacement $\chi_1 \rightarrow \Phi~~;~~m_1 \rightarrow M~~;~~\chi_2 \rightarrow \chi~~;~~ m_2 \rightarrow 0$.

In all the cases studied in this article, namely particle decay and those that feature threshold and infrared singularities, the initial single particle state ``decays'' either exponentially or with a power law and is not an asymptotic state. The asymptotic states that result from the time evolution of these processes are given by  (\ref{comonasy},\ref{comonasyfichi}), these  are  correlated  kinematically entangled states  of the daughter particles.  We highlight this noteworthy point:  particle decay and the processes that feature threshold and infrared divergences, while quantitatively different in the details of the dynamical evolution of the amplitudes,  are asymptotically qualitatively similar and determined by the asymptotic states (\ref{comonasy},\ref{comonasyfichi}) which  characterize  the \emph{production} of the daughter particles, with  the total production probability fulfilling the unitarity relation, namely $\sum_{\vp}|C^{\chi}_{\vp;\vk}(\infty)|^2 =1$.  Hence, in conclusion,   threshold and infrared singularities result in the production of the daughter particles, much in the same manner as the usual decay process.

Out of the pure state (\ref{comonasy}) (or (\ref{comonasyfichi})), we can construct the (pure state) density matrix
\be \varrho = |\Psi_I(\infty)\rangle\langle \Psi_I(\infty)| ~~;~~ \mathrm{Tr}\varrho = 1 \,,\label{densmtx}\ee  where the identity in the trace is a result of  the unitarity relation (\ref{asysr}). Consider taking expectation values of operators that act on the Hilbert space of only one of the fields, for example an operator $\mathcal{O}^{(\chi_1)} $ that acts solely on the Hilbert space of the field $\chi_1$
\be \langle \mathcal{O}^{(\chi_1)} \rangle \equiv \mathrm{Tr}_{\chi_1,\chi_2}\big[\varrho\,\mathcal{O}^{(\chi_1)}\big] \,,\label{tracechi1} \ee or similarly, of operators that act solely on the Hilbert space of the field $\chi_2$. In these cases the trace over the ``unobserved'' fields yields a reduced density matrix, namely
\be \varrho_{\chi_1} = \mathrm{Tr}_{\chi_2} \varrho ~~;~~ \varrho_{\chi_2} = \mathrm{Tr}_{\chi_1} \varrho \,,\label{redrhos}\ee
and the unitarity condition obviously yields
\be \mathrm{Tr}_{\chi_1} \varrho_{\chi_1} = 1 ~~;~~ \mathrm{Tr}_{\chi_2} \varrho_{\chi_2} = 1 \,. \label{tracesrhos}\ee From the asymptotic state (\ref{comonasy}) we find
\be \varrho_{\chi_1} = \sum_{\vp} |C^{\chi}_{\vp;\vk}(\infty)|^2\,|1^{\chi_1}_{\vp}\rangle \langle 1^{\chi_1}_{\vp}| \,,\label{varho1} \ee
\be \varrho_{\chi_2} = \sum_{\vp} |C^{\chi}_{\vp;\vk}(\infty)|^2\,|1^{\chi_2}_{\vk-\vp}\rangle \langle 1^{\chi_2}_{\vk-\vp}| \,.\label{varho2} \ee
These reduced density matrices describe \emph{mixed states}, and are diagonal in momentum and particle number. In particular we identify the
distribution function of the produced particles in terms of the expectation value of the number operator for each field $\mathcal{N}_{\chi}(\vp)$ as
\be  \langle \mathcal{N}_{\chi_1}(\vp) \rangle = \mathrm{Tr}_{\chi_1} \mathcal{N}_{\chi_1}(\vp) \, \varrho_{\chi_1} = |C^{\chi}_{\vp;\vk}(\infty)|^2 \,, \label{number1}\ee
\be  \langle \mathcal{N}_{\chi_2}(\vq) \rangle = \mathrm{Tr}_{\chi_2} \mathcal{N}_{\chi_2}(\vq) \, \varrho_{\chi_2} = |C^{\chi}_{\vk-\vq;\vk}(\infty)|^2 \,, \label{number2}\ee therefore, as a consequence of entanglement, both daughter particles share the same distribution function. Furthermore, as a consequence of unitarity   we find
\be \sum_{\vp}\langle \mathcal{N}_{\chi_1}(\vp) \rangle = \sum_{\vq} \langle \mathcal{N}_{\chi_2}(\vq) \rangle =1\,, \label{totalnum}\ee
a result with the clear interpretation that there are in total one $\chi_1$ and one $\chi_2$ particles in the asymptotic state, a physically correct outcome of the generalized decay of a single $\Phi$ particle into one $\chi_1$ and  one $\chi_2$ particles.

The results obtained above  allow us to obtain the distribution function in the cases under consideration from the general expression
(\ref{finprobchi2asy}). Denoting the matrix element squared in (\ref{finprobchi2asy}) by $\mathcal{M}_{\vp,\vk}$ we find for the
case of decay, namely $E^\Phi_k > k_{0T}$
\be |C^{\chi}_{\vp;\vk}(\infty)|^2 =  \, \frac{ 2\, \mathcal{M}_{\vp,\vk} \,\mathcal{Z}_d }{\big[ E^{\chi_1}_{\vp}+E^{\chi_2}_{\vq}-E^{\Phi}_{k} \big]^2+  \Gamma^2_k  }  \,, \label{distdecay} \ee
which    can be written
 as
\be  |C^{\chi}_{\vp;\vk}(\infty)|^2 = \frac{2\pi}{\Gamma_k}\, \mathcal{M}_{\vp,\vk}\,\mathcal{Z}_d \, \frac{1}{\pi}\, \frac{  \Gamma_k   }{\big[E^{\Phi}_{k}- E^{\chi_1}_{\vp}-E^{\chi_2}_{\vq} \big]^2+  \Gamma^2_k  }\,, \label{loren} \ee the replacement (\ref{narwid}) in the narrow width limit yields a sharp  energy conserving delta function, however, a small but finite width introduces an energy uncertainty in the distribution of daughter particles as a consequence of the lifetime $1/\Gamma_k$ of the initial state with a concomitant broadening of the distribution function.

For the case of infrared singularity, namely $m_1=M, m_2 =0$ we find
\be |C^{\chi}_{\vp;\vk}(\infty)|^2 =  \, \frac{ 2\,\mathcal{M}_{\vp,\vk}  }{\big[E^{\Phi}_{k}- E^{\chi_1}_{\vp}-E^{\chi_2}_{\vq} \big]^2  } ~\Bigg[ \frac{[ E^{\chi_1}_{\vp}+E^{\chi_2}_{\vq}-E^{\Phi}_{k} }{E^{\Phi}_{k}}\Bigg]^{2g^2}  \,. \label{distinfra} \ee
This distribution function does not feature any particular scale, although it is peaked at $\Omega= E^{\chi_1}_{\vp}+E^{\chi_2}_{\vq}-E^{\Phi}_{k} =0$, namely energy conserving transitions, it falls off as a power law  $ {\Omega^{-2(1-g^2)}}$ consistently with the scale invariance and anomalous dimension associated with infrared phenomena found in ref.\cite{infra} .

\begin{figure}[h!]
\begin{center}
\includegraphics[height=4in,width=4in,keepaspectratio=true]{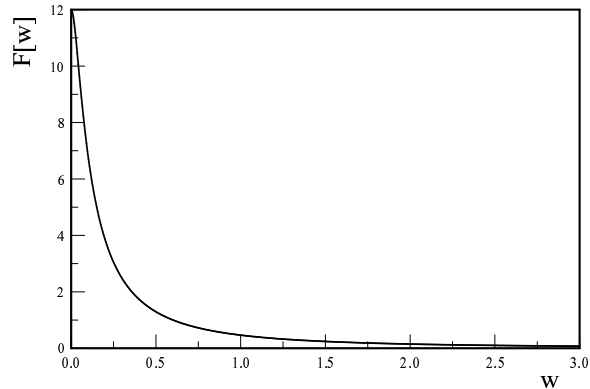}
\caption{The function $F[w]$ vs $w$.}
\label{fig10:fofw}
\end{center}
\end{figure}

For the case of threshold singularity  we can write the distribution function as
\be |C^{\chi}_{\vp;\vk}(\infty)|^2 =    2\,(t^*)^2 \,\mathcal{M}_{\vp,\vk}\, \mathcal{Z}_{ir}\,{F}[w]  ~~;~~ w= \big[E^{\chi_1}_{\vp}+E^{\chi_2}_{\vq}-E^{\Phi}_{k}  \big]\,t^* ~~;~~ t^* = \frac{1}{4\pi\,g^4\,M}\,\frac{E^\Phi_k}{M}\,,  \label{distthres}\ee
where
\be  {F}[w] = \int^\infty_0 \frac{\sin[w\,\tau]}{w} \, e^{-\sqrt{\tau}}\,d\tau   \,, \label{funcofwt}  \ee although there is an analytic expression for this integral in terms of Fresnel integral functions,  a graphical representation is more  illuminating and is displayed in fig. (\ref{fig10:fofw}). The distribution function    is sharply peaked at $\Omega = \big[E^{\chi_1}_{\vp}+E^{\chi_2}_{\vq}-E^{\Phi}_{k}  \big] \simeq  0$  with a width of the order of $1/t^{*}$ consistent with the ``lifetime'' of the initial state. Note that this distribution is \emph{narrower} than the case of decay because the lifetime $t^* \propto 1/g^4$ is longer as compared with $1/\Gamma \propto 1/g^2$.

The reduced density matrices (\ref{varho1},\ref{varho2}) describe mixed states, therefore there is an associated von Neumann entropy with each, this is the entanglement entropy arising from the loss of information as a result of tracing over the complementary degree of freedom.
This would be the case for example if one of the particles in the decay process belongs to a dark sector beyond the standard model and is not observable, as in an ``invisible decay''.

 Because the pure state (\ref{comonasy}) is an entangled state of the $\chi_1,\chi_2$ particles both subsystems share the same entanglement entropy
\be S_{vN} = -\sum_{\vp} |C^{\chi}_{\vp;\vk}(\infty)|^2\,\ln\Big[|C^{\chi}_{\vp;\vk}(\infty)|^2 \Big] >0 \,.\label{entropy}\ee Since initially the coefficients vanish, the entanglement entropy vanishes, whereas it is positive asymptotically at long time since by unitarity $|C^{\chi}_{\vp;\vk}(\infty)|^2<1$. Therefore the entanglement entropy $S_{vN}$ grows as a consequence of  unitary time evolution, its time evolution is completely determined by the dynamical resummation equations (\ref{Ckapasol},\ref{intdiff}).

The entanglement entropy has also been discussed in the case of decay in ref.\cite{entdecay}, and for infrared singularity within the context of quantum electrodynamics in refs.\cite{infra,infrafrw,tomaras}. While the entanglement entropy is a corollary of the pair correlation in the asymptotic state, it is just beginning to receive attention within particle physics\cite{kharzeev}.

\section{Discussion}\label{sec:discussion}

\vspace{1mm}

\textbf{General lessons: common aspects of decay, threshold and infrared divergences: } Although we have focused our study on a simple quantum field theory,
the results obtained in the previous section suggest some universality in the asymptotic state arising from decay, threshold or infrared singularities in that in all these cases the asymptotic state is a kinematically entangled multiparticle state with a probability that saturates the unitarity constraint. While this is obviously a consequence of unitary time evolution, the corollary is that threshold and infrared divergences are just as efficient mechanisms of particle production as the process of decay. The asymptotic distribution functions
$|C^{\chi}(\infty)|^2$ are peaked at $\Omega = E^{\chi_1}_{\vp}+E^{\chi_2}_{\vq}-E^{\Phi}_{k} =0$, namely energy conserving transitions, but broadened. In the case of decay the width of this distribution is $\mathcal{O}(\Gamma) \propto g^2 $ consistent with a broadening by the lifetime $\propto 1/\Gamma$,  for threshold singularity the distribution is \emph{narrower}, within a width of $\mathcal{O}(1/t^*) \propto g^4$ again consistent with a much longer ``lifetime'', and in the case of infrared singularity the distribution function features a scaling behavior with anomalous dimension $\Omega^{-2(1-g^2)}$, as a consequence of the scale invariance and anomalous dimension associated with infrared phenomena\cite{infra}.

The detailed analysis of the different cases yield the following set of criteria on the spectral density  $\rho(k_0)$ and the mass of the particle  that determines  the time evolution of the survival probabilities:

\textbf{a:} If the spectral density does not vanish at $k_0 = E_k$ where $E_k$ is the single particle energy of the ``decaying field'', the survival probability decays as usual $\propto e^{-\Gamma t}$ with $\Gamma = 2\pi \rho(E_k)$. This is simply the statement of Fermi's golden rule and is the S-matrix result for the decay width at leading order in the coupling.

\textbf{b:} If $E_k$ coincides with the multiparticle threshold and the spectral density vanishes at threshold as a square root $\propto \sqrt{|k_0-k_{0T}|}$ this case corresponds to a threshold singularity. The usual decay rate vanishes but the survival probability decays as $e^{-\sqrt{t/t^*}}$. This result cannot be obtained within the S-matrix approach, since the transition probability per unit time in the infinite time limit, namely the usual decay rate calculated via S-matrix vanishes.

\textbf{c:} If $E_k$ coincides with the multiparticle threshold and the spectral density vanishes linearly at threshold  $\propto |k_0-k_{0T}|$ this case corresponds to an infrared singularity. The usual decay rate vanishes but the survival probability decays algebraically with an anomalous dimension $t^{-\Delta}$. Again, this result cannot be obtained via the usual S-matrix calculation for the transition probability per unit time in the infinite time limit, again such decay rate vanishes.

More generally, if the spectral density vanishes at threshold as $|k_0-k_{0T}|^\beta$ the survival probability decays as $e^{-C\,t^{1-\beta}}$ with $C$ a
coupling dependent constant, $\beta =1$ is the infrared singular case and yields a logarithmic behavior.

Thus,   threshold and infrared singularities  differ only  on how the spectral density vanishes at threshold: if as a square root, then the decay is $e^{-\sqrt{t/t^*}}$ yielding a distribution function with a breadth $\propto 1/t^*$, if linearly, the decay is $\propto t^{-\Delta}$  yielding a distribution function with scaling dimension $2-\Delta$.

\vspace{1mm}

\textbf{Infrared and threshold divergences as production mechanism of ultralight particles:} Although we have studied the dynamics associated with infrared divergences for the case in which the $\chi_2$ particle is massless, the results apply to the case of ultralight particles proposed to be dark matter candidates, from ``fuzzy'' dark matter with a mass $\simeq 10^{-22} \,\mathrm{eV}$\cite{fuzzyDM,fuzzy2,wittenfuzzy}, to axions with a mass $\simeq 10^{-6}\,\mathrm{eV}$\cite{axionreviu,axionsikivie}.  Consider that such particles are coupled to heavier one, with a mass    $\gtrsim 100 \mathrm{MeV}$, the departure from threshold is $\lesssim 10^{-14}$ of the value of the threshold position, this means that  although the threshold is just above the single particle pole, the wave function renormalization is vanishingly small (see figures (\ref{fig4:zcabove},\ref{fig6:zetair})) thus transferring the normalization to the continuum, and the
  time evolution --either as a square root or logarithmic-- lasts for a very long time thus populating the asymptotic state with the ultralight degree of freedom. Thus infrared or threshold divergences are an efficient mechanism for production of ultralight dark matter candidates as proposed recently in ref.\cite{infrafrw}.

\vspace{1mm}

\textbf{Fermion loops:} Threshold divergences depend crucially on the behavior of the spectral density at threshold. Whereas for the case of a bosonic loop the spectral density near threshold vanishes as $(1-4m^2/M^2)^{1/2}$ yielding the $\gamma(t) \rightarrow \sqrt{t}$ and the decay law (\ref{thresdec}), a fermion loop yields a spectral density that vanishes as $(1-4m^2/M^2)^{3/2}$ yielding $\gamma(t) \rightarrow 1/\sqrt{t}$ thus approaching an (ultraviolet divergent) constant at long time. The lack of a threshold divergence in the case of a fermion loop has also been recognized in ref.\cite{will}.

\vspace{1mm}

\textbf{Relaxation and thermalization:} Both for threshold and infrared divergences the usual decay rate vanishes, however the
survival probability decays either as $e^{-\sqrt{t/t^*}}$ or as $t^{-2g^2}$, in either case the decay law cannnot be described by  Fermi's golden rule or the S-matrix approach. This observation leads to the question of how $\Phi$ particles would thermalize with a bath of $\chi$ particles under the conditions of threshold or infrared divergence. In the usual Boltzmann equation, the thermalization rate is directly proportional to the decay rate obtained from Fermi's golden rule modified by spontaneous emission/absorption factors. This question is of relevance in cosmology and requires a treatment different from the Boltzmann equation which directly inputs the transition probabilities per unit time from S-matrix theory, these are precisely the relaxation rates from Fermi's golden rule which vanish for threshold or infrared divergences. A related question is how detailed balance emerges between decay and inverse decay processes, since in the usual formulation detailed balance is a consequence of explicit energy conservation and the energy conserving constraint is not exactly satisfied for threshold and infrared divergences, this is the reason that the usual decay rate vanishes in these cases.  Work on these aspects is in progress and will be reported elsewhere\cite{boyatherma}.

\textbf{Entanglement entropy, correlations and thermalization}

We have discussed the emergence of the entanglement entropy  upon tracing  an ``unobserved'' degree of freedom out of the pure asymptotic state density matrix. Such tracing, or coarse  graining, yields a mixed state, and a concomitant von Neumann entropy as a consequence of the loss of information in the coarse graining process. The pair correlations in the pure entangled state entail that the reduced density matrices $\varrho_{\chi_1},\varrho_{\chi_2}$   feature the same probabilities, (see eqns.\ref{varho1},\ref{varho2}) which are   identified as the distribution function of the produced particles, hence the same entanglement entropy.

A remarkable experiment reported in ref.\cite{entantherma}) shows that the entanglement entropy as a result of correlations in a closed system heralds thermalization. It is therefore an intriguing possibility that in the early Universe, indeed a closed system, the entanglement entropy associated with cosmological particle production from threshold or infrared divergences\cite{infrafrw} may also herald the onset of a thermal state.

Entanglement plays a fundamental role in the determination of time reversal and CP violation in neutral meson systems\cite{meson}, therefore it is a tantalizing possibility that correlations of particles produced via threshold or infrared divergences may prove to be also relevant in experimental particle physics. The potential relevance of the concept of entanglement entropy associated with information loss in the asymptotic final state, in particular if some of the decay products belong to a dark sector beyond the standard model, both in cosmology and in particle physics   merits further study.

\textbf{Phenomenological consequences of the lifetime for threshold divergences:}  The generalized decay as a consequence of threshold divergences with a survival probability that decays as $e^{-\sqrt{t/t^*}}$ implies that even when the usual decay rate vanishes (infinite lifetime), there is an intrinsic finite lifetime $t^* \propto 1/g^4$. This result may have potentially relevant phenomenological implications, as the decay products of this process may feature displaced vertices with a very long but finite decay length.

\vspace{1mm}

\textbf{Radiative corrections: moving away from threshold.} The condition for threshold divergence, namely that the mass of the particle coincides exactly with the value of the lowest multiparticle threshold, will most likely not survive radiative corrections. However, such corrections will be proportional to a power of a small coupling, thus while not exactly at threshold, the departure from threshold is perturbatively small. Let us consider that upon radiative corrections the mass of the particle moves perturbatively below threshold so that $ (4m^2-M^2)/M^2 \propto \alpha$ with $\alpha$ a small coupling. In this case the particle has been rendered stable by radiative corrections, however, asymptotically its probability in the final state is $\mathcal{Z} \propto e^{-\frac{c}{\sqrt{\alpha}}}$ with $c$ a constant of $\mathcal{O}(1)$, hence featuring an essential singularity in the coupling and for all intent and purpose the particle does not appear as an asymptotic state. If, on the other hand, the radiative correction moves the mass above threshold, the particle is unstable, decaying as $e^{-\sqrt{t/t^*}} $ during a time $\propto 1/\alpha$ until it begins decaying as $e^{-\Gamma t}$, and is not an asymptotic state. In conclusion, radiative corrections while capable of moving the position of the mass shell away from threshold perturbatively, the probability of the particle to be present in the asymptotic state practically vanishes.

 For the case of infrared divergence, for example for the model defined by the Lagrangian density (\ref{charged}) in which a massive charged particle emits and absorbs a massless $\chi$ particle, unless the mass of this particle is protected by some symmetry  radiative corrections  will induce a non-vanishing mass thus modifying the conclusions. However if such modification is perturbatively small, the mass shell of the charged particle will be very close to threshold and the near-threshold behavior will ensue as discussed above.

\vspace{1mm}

\section{Conclusions}

Motivated by the possibility that a dark sector beyond the standard model could feature ultralight particles as dark matter candidates, in this article we study  threshold and infrared divergences as hitherto unexplored possible production mechanisms that could be relevant in cosmology. In the case of threshold and infrared divergences the usual decay rates vanish, therefore understanding the time evolution in these cases will pave the way towards understanding the process of thermalization beyond the usual Boltzmann approach which inputs the transition rates per unit time in the infinite time limit.  Our main objectives are to compare the usual decay process  to the time evolution and particle production associated with threshold and infrared divergences and to understand the nature and characteristics of the asymptotic state. We study these different mechanisms in a model field theory that provides a simple arena to explore these phenomena within the same setting by varying the masses of the various fields yet allows to extract more general lessons. An analysis based on the Kallen-Lehmann representation of the  particle propagator suggests that decay, threshold and infrared singularities, while seemingly widely different phenomena are qualitatively related, and also highlights   the breakdown of a Breit Wigner approximation to propagators in the cases of threshold and infrared divergences. A dynamical resummation method complementary to the dynamical renormalization group is introduced to study the time evolution of initially prepared single particle states. This method is manifestly unitary and yields the asymptotic state, from which we obtain the \emph{distribution function} of the produced particles. We find that whereas in a typical decay process the survival probability of the initial single particle state decays as $e^{-\Gamma t}$,  in the case of threshold divergence it decays as $e^{-\sqrt{t/t^*}}$ and for the case of infrared divergence $t^{-\Delta}$, where $\Gamma$ and $\Delta$ are $\propto (coupling)^2$ while $t^* \propto 1/(coupling)^4$. Although the decay laws are strikingly different, the asymptotic state is more ``universal'' in the sense that it is a kinematically entangled state of the daughter particles. The probability of the asymptotic state is shown in each case to satisfy the unitarity condition. The distribution function of the particles in the asymptotic state are strongly peaked at energy conserving transitions, but in the case of the usual decay and of threshold singularity they are broadened by the lifetime of the
decaying state $1/\Gamma$, $t^*$ respectively, whereas in the case of the infrared divergence the distribution function falls off with a scaling behavior with scaling dimension $2-\Delta$.

Therefore the results of this study indicate that threshold and infrared divergences are production mechanisms just as efficient as the usual particle decay.
If either one of the particles in the final state is not  observed  as perhaps in an ``invisible decay'' into a dark matter particle, the information loss leads to an entanglement entropy which grows as a consequence of unitary time evolution.
These alternative mechanisms may be relevant for production of particles in the dark sector in cosmology with possible phenomenological consequences in invisible decays with displaced vertices and long decay lengths, and also to novel thermalization dynamics, a possibility that merits further study.

\acknowledgements
  The author gratefully acknowledges support from the U.S. National Science Foundation through grant award NSF 2111743.

 \appendix

 \section{Spectral density.}\label{app:sd}
 Upon quantization in a volume $V$ in  a discrete momentum representation, the relevant matrix element for the interaction described by the interaction Hamiltonian in the interaction picture (\ref{Hint}) is found to be
 \be \bra{1^{\Phi}_{\vk}}H_I(0)\ket{1^{\chi_1}_{\vp};1^{\chi_2}_{\vq}} = \frac{\lambda}{V^{1/2}}\, \frac{\delta_{\vp+\vq,\vk}}{\big[8\,E^{\Phi}_k\,E^{\chi_1}_{\vp}\,E^{\chi_2}_{\vq}\big]^{1/2}}\,. \label{mtxele}\ee

 This matrix element makes explicit  momentum conservation.

 The spectral density is defined by eqn. (\ref{rhopo}), with the matrix elements given by eqn. (\ref{mtxele}) and  passing to the continuum limit with $\frac{1}{V} \sum_{\vp} \rightarrow \int \frac{d^3p}{(2\pi)^3}$ we recognize that (\ref{rhopo}) is given by
 \be \rho_{\Phi}(k_0) = \frac{\lambda^2}{8\,E^{\Phi}_{\vk}}\,\int \frac{\delta\Big(k_0-E^{\chi_1}_{\vp}-E^{\chi_2}_{\vk-\vp}\Big) }{E^{\chi_1}_{\vp}\,E^{\chi_2}_{\vk-\vp}}\,\frac{d^3p}{(2\pi)^3}\,,\ee
 which is the Lorentz invariant two body phase space, finally yielding
 \be \rho_{\Phi}(k_0) = g^2  \,\frac{M^2}{2E^{\Phi}_k}\,\Bigg[\Big(1-\frac{(m_1+m_2)^2}{k^2_0-k^2} \Big)\Big(1-\frac{(m_1-m_2)^2}{k^2_0-k^2} \Big) \Bigg]^{1/2}\,\Theta(k^2_0-k^2-(m_1+m_2)^2)\,\Theta(k_0)\,. \label{rhofina}\ee To leading order in the coupling we   replaced $\Big( \frac{\lambda}{4\pi \, M}\Big)^2 \rightarrow g^2$ where $g$ is the dimensionless coupling introduced in eqn. (\ref{dimratios}).

 \section{Long time limit of $\gamma(t)$ in decay case:}\label{app:longama}
 For the case of decay, $\gamma(t)$ is given by eqn. (\ref{gamares}) with $\overline{\rho}(x/Mt)$ given by eqn. (\ref{overho}). Let us write  eqn. (\ref{gamares}) as
 \be \gamma(t) = \frac{g^2 M}{E^{\Phi}_k}\,(Mt) \,I(t) \ee where
 \be I(t) =  \underbrace{\int^{X(t)}_{-X(t)} \overline{\rho}(x/Mt)\,\frac{1-\cos(x)}{x^2}\, dx}_{I_1(t)} +  \underbrace{\int^{\infty}_{X(t)} \overline{\rho}(x/Mt)\,\frac{1-\cos(x)}{x^2} \, dx }_{I_2(t)}\,, \label{Iint} \ee in $I_1(t)$ the integration interval is symmetric and the function $(1-cos(x))/x^2$ is even in $x$, therefore only the symmetric combination   $\overline{\rho}_e(x/Mt)= (\overline{\rho}(x/Mt)+ \overline{\rho}(-x/Mt))/2 $ contributes to this integral. Adding and subtracting $\overline{\rho}(0)$, it follows that
 \be I_1(t) = \overline{\rho}(0)\,\int^{X(t)}_{-X(t)}  \frac{1-\cos(x)}{x^2} \, dx  + \int^{X(t)}_{-X(t)} \Big[\overline{\rho}_e(x/Mt)- \overline{\rho}(0) \Big]\,\frac{1-\cos(x)}{x^2} \, dx \, \label{I1v2}\ee
\be  \int^{X(t)}_{-X(t)}  \frac{1-\cos(x)}{x^2} \, dx = 2\,\Bigg\{\underbrace{\int^{\infty}_{0}  \frac{1-\cos(x)}{x^2}\,dx}_{\frac{\pi}{2}} - \,\int^{\infty}_{X(t)}  \frac{1-\cos(x)}{x^2} \, dx \Bigg\}\,, \label{secin} \ee in the second integral in (\ref{secin}) change variables to
\be x = \xi M t,  \Rightarrow X(t) = \overline{\xi} Mt~~;~~ \overline{\xi} = (E^\Phi_k-k_{0T})/M >0 \label{nuvars}\ee in terms of which this second integral becomes
\be \frac{1}{Mt}\, \int^{\infty}_{\overline{\xi}}  \frac{1-\cos(\xi \,Mt)}{\xi^2} \, d\xi ~~ {}_{\overrightarrow{Mt \rightarrow \infty} } ~~~ \frac{1}{Mt\, \overline{\xi}} \ee where the cosine term averages out in the long time limit (Riemann-Lebesgue lemma). Performing the same change of variables in the second integral in (\ref{I1v2}), yields for this contribution
\be \frac{1}{Mt}\, \int^{\overline{\xi}}_{-\overline{\xi}} \Big[\overline{\rho}_e(\xi)- \overline{\rho}(0) \Big]\,\frac{1-\cos(\xi Mt)}{\xi^2} \, d\xi \ee because $\overline{\rho}_e$ is even in $\xi$ it follows that for $\xi \simeq 0$ the numerator is of $\mathcal{O}(\xi^2)$, therefore cancelling the $\xi^2$ in the denominator, hence the region of integration near the origin yields
a vanishing contribution and the cosine term oscillates averaging out in the $Mt\rightarrow \infty $ limit. In this limit the second integral in (\ref{I1v2}) yields
\be \frac{1}{Mt}\, \int^{\overline{\xi}}_{-\overline{\xi}} \Big[\overline{\rho}_e(\xi)- \overline{\rho}(0) \Big]\,    \frac{d\xi}{\xi^2} \,. \ee Gathering all the terms we find
\be I_1(t) ~~ {}_{\overrightarrow{Mt \rightarrow \infty}} ~~ \pi\,\overline{\rho}(0) - \frac{2\,\overline{\rho}(0)}{Mt\, \overline{\xi}}\,+ \frac{1}{Mt}\, \int^{\overline{\xi}}_{-\overline{\xi}} \Big[\overline{\rho}_e(\xi)- \overline{\rho}(0) \Big]\,    \frac{d\xi}{\xi^2} \ee

Carrying out the same change of variables in $I_2(t)$ in eqn. (\ref{Iint}) and taking the long time limit $Mt \rightarrow \infty$ in which the cosine term averages out as in the previous integrals, yielding
\be I_2(t) ~~{}_{\overrightarrow{Mt\rightarrow \infty}} ~~ \frac{1}{Mt} \, \int^{\infty}_{\overline{\xi}} \frac{\overline{\rho}(\xi)}{\xi^2}\,     {d\xi} \,. \ee Including all contributions, we finally find in the long time limit
\be \gamma(t) = \Gamma_k t + 2 z_d \label{gamafinale}\ee where
\be \Gamma_k = \frac{\pi\,g^2 M^2 }{E^{\Phi}_k}\,\overline{\rho}(0) = \frac{\pi\,g^2 M^2 }{E^{\Phi}_k}\,\sqrt{1-\frac{4m^2}{M^2}} \label{gamapp} \,,\ee and
\be 2z_d = \frac{ g^2 M }{E^{\Phi}_k}\,\Bigg\{-\frac{2\,\overline{\rho}(0)}{  \overline{\xi}}\,+  \, \int^{\overline{\xi}}_{-\overline{\xi}} \Big[\overline{\rho}_e(\xi)- \overline{\rho}(0) \Big]\,    \frac{d\xi}{\xi^2} + \int^{\infty}_{\overline{\xi}} \frac{\overline{\rho}(\xi)}{\xi^2}\,    {d\xi}  \Bigg\}~~;~~ \overline{\xi} = (E^{\Phi}_k-k_{0T})/M \,.  \label{zdapp} \ee

\section{Useful identity:}\label{app:identity}

 From eqn. (\ref{Ckapasol}) we find

\be  C^{\chi}_{\vp;\vk}(t)   =    -i \,\langle 1^{\chi_1}_{\vp};1^{\chi_2}_{\vq} |H_I(0)|1^{\Phi}_{\vk}\rangle \int_0^t e^{ i\Omega t'}\, \,C^{\Phi}_{\vk}(t')\,dt' ~~;~~ \Omega \equiv  E^{\chi_1}_{\vp}+E^{\chi_2}_{\vq} - E^{\Phi}_k \,, \label{Chis2}  \ee  hence

\be |C^{\chi}_{\vp;\vk}(t)|^2 = |\langle 1^{\chi_1}_{\vp};1^{\chi_2}_{\vq} |H_I(0)|1^{\Phi}_{\vk}\rangle|^2 \int^t_0 \int^t_0  e^{i\Omega t_1} \, C^{\Phi}_{\vk}(t_1)\,e^{-i\Omega t_2} \, (C^{\Phi}_{\vk}(t_2))^*\,dt_1\,dt_2 \,. \label{modC2}\ee

Inside the integrals we replace the amplitudes $ C^{\Phi}_{\vk}(t)$ by eqn. (\ref{solumarkov}). Since at early time  the amplitude departs from   $C^{\Phi}_{\vk}(0) =1$ by a perturbatively small amount, we will replace them by their long time limit

  \be  C^{\Phi}_{\vk}(t) = e^{-i\delta E_{\infty}\,t}\, e^{-\frac{\gamma(t)}{2}}  \,,\label{ampslt}\ee where $\gamma(t)$ is taken in the long time limit for  the different cases,   and absorb $\delta E_{\infty}$  into a renormalization of $E^{\Phi}_k$.

  The integrand in the double time integral in  (\ref{modC2}) is now given by ($E^{\Phi}_k$ in $\Omega$ now stands for the renormalized energy)
\be Q(t_1,t_2) = e^{i\,\Omega(t_1-t_2)}\,e^{-\frac{1}{2}(\gamma(t_1)+\gamma(t_2))}\,, \label{J12}\ee writing the double time integral in (\ref{modC2}) as
\be \int^t_0 \int^t_0 Q(t_1,t_2) \,\Big(\Theta(t_1-t_2)+\Theta(t_2-t_1)\Big)\,dt_1\,dt_2 = 2 \int^t_0 dt_1 e^{-\frac{\gamma(t_1)}{2}}\int^{t_1}_0 \cos[\Omega(t_1-t_2)]\,e^{-\frac{\gamma(t_2)}{2}} \,   dt_2 \,,\label{dfun}\ee where in the term with $\Theta(t_2-t_1)$ on the left hand side of (\ref{dfun}) we relabelled $t_1 \leftrightarrow t_2$ and used that $Q(t_2,t_1) = Q^*(t_1,t_2)$ with $\gamma(t)$ being real. Now writing
\be \cos[\Omega(t_1-t_2)] = \frac{d}{dt_2}   \int^{t_2}_0 \cos[\Omega(t_1-t')]\,dt'  \,,\label{Gf}\ee in the $t_2$ integral in (\ref{dfun}), we integrate by parts using (\ref{Gf}) and neglect  the term proportional to the time derivative of $\gamma(t_2)$ because it is
of $\mathcal{O}(H^2_I)$, and because the modulus squared of the matrix element in (\ref{modC2}) is of order $H^2_I$, neglecting the derivative of $\gamma$ is consistent with neglecting   terms of $\mathcal{O}(H^4_I)$ in (\ref{modC2}). Therefore, up to $\mathcal{O}(H^4_I)$  we find that the double integral in (\ref{modC2}) becomes
\be \int^t_0 \int^t_0  e^{-i\Omega t_1} \,  C^{\Phi}_{\vk}(t_1)\,e^{i\Omega t_2} \,  (C^{\Phi}_{\vk}(t_2))^*\,dt_1\,dt_2 = \frac{2}{\Omega} \,\int^t_0 \sin\big[\Omega\,t_1\big]\,e^{-\gamma(t_1)}\, dt_1\,. \label{findoubint}\ee Inserting this result into eqn. (\ref{modC2}) we find the final expression for the probabilities valid up to $\mathcal{O}(H^4_I)$,
\be |C^{\chi}_{\vp;\vk}(t)|^2 = \frac{2}{\Omega}\,|\langle 1^{\chi_1}_{\vp};1^{\chi_2}_{\vq} |H_I(0)|1^{\Phi}_{\vk}\rangle|^2 \, \int^t_0 \sin\big[\Omega\,t_1\big]\,e^{-\gamma(t_1)}\, dt_1 ~~;~~ \Omega \equiv  E^{\chi_1}_{\vp}+E^{\chi_2}_{\vq} - E^{\Phi}_k \,, \label{finprobchi2}\ee
and $E^\Phi_k$ here is the renormalized single $\Phi$ particle energy.

\end{document}